\documentclass[aps,prd,showpacs,floatfix,superscriptaddress]{revtex4-2}
\usepackage{graphicx}
\usepackage{subcaption}
\usepackage{dsfont}
\usepackage{xcolor}
\usepackage{multirow}
\usepackage[normalem]{ulem}
\usepackage{enumitem} 
\usepackage{lineno}
\usepackage{bm}
\usepackage{comment,slashed}
\definecolor{refcol}{RGB}{34,34,178}%{178,34,34}%{0,100,170}%{0,0,205}
\usepackage[colorlinks,linkcolor=blue,citecolor=blue,urlcolor=blue]{hyperref}
\usepackage{microtype}
\usepackage{float}
\usepackage{appendix}
\usepackage{soul}
\usepackage{amsmath,amssymb}
\graphicspath{{./Fig/}}

\newcommand{\RNum}[1]{\uppercase\expandafter{\romannumeral #1\relax}}

\begin{document}

\title{QCD phase diagram in the $T-eB$ plane for varying pion mass}

\date{ \today }

\author{Mahammad Sabir Ali}
\email{sabir@niser.ac.in}
\affiliation{School of Physical Sciences, National Institute of Science Education and Research, An OCC of Homi Bhabha National Institute, Jatni-752050, India}
\affiliation{Department of Theoretical Physics, Tata Institute of Fundamental Research, Mumbai-400005, India}

\author{Chowdhury Aminul Islam}
\email{chowdhury@physik.uni-frankfurt.de}
\affiliation{Institut f\"{u}r Theoretische Physik, Johann Wolfgang Goethe–Universit\"{a}t, Max-von-Laue-Str. 1, D–60438 Frankfurt am Main, Germany}
\affiliation{School of Nuclear Science and Technology, University of Chinese Academy of Sciences, Beijing 100049, China}

\author{Rishi Sharma}
\email{rishi@theory.tifr.res.in}
\affiliation{Department of Theoretical Physics, Tata Institute of Fundamental Research, Mumbai-400005, India}

%%%%%%%%%%%%%%%%%%%%%%%%%%%%%%%%%%%%%%%%%%%%%%%%%%%%%%%%%%%%%%%%%%%%%%%%

\begin{abstract}  
We study the effect of a varying pion mass on the quantum chromodynamics (QCD) phase diagram in the presence of an external magnetic field, aiming to understand it, for the first time, using Nambu\textendash Jona-Lasinio like effective models. We compare results from both its local and nonlocal versions. In both cases, we find that the inverse magnetic catalysis (IMC) near the crossover is eliminated with increasing pion mass, while the decreasing trend of crossover temperature with increasing magnetic field persists for pion mass values at least up to $440$ MeV. Thus, the models are capable of capturing qualitatively the results found by lattice QCD (LQCD) for heavy (unphysical) pions. The key feature in the models is the incorporation of the effect of a reduction in the coupling constant with increasing energy. Along with reproducing the IMC effect, it enables models to describe the effects of heavier current quark masses without introducing additional parameters. For the local NJL model, this agreement depends on how the parameters of the model are fit at the physical point. In this respect, the nonlocal version, which, due to its formulation, automatically exhibits the IMC effect around the crossover region, captures the physics more naturally. We further use the nonlocal framework to determine the pion mass beyond which the IMC effect around the transition region does not exist anymore. 
\end{abstract}

%%%%%%%%%%%%%%%%%%%%%%%%%%%%%%%%%%%%%%%%%%%%%%%%%%%%%%%%%%%%%%%%%%%%%%%%

\maketitle

%\linenumbers
\clearpage

%%%%%%%%%%%%%%%%%%%%%%%%%%%%%%%%%%%%%%%%%%%%%%%%%%%%%%%%%%%%%%%%%%%%%%%%
\section{Introduction}
\label{sec:introduction}
%%%%%%%%%%%%%%%%%%%%%%%%%%%%%%%%%%%%%%%%%%%%%%%%%%%%%%%%%%%%%%%%%%%%%%%%
The effect of an external magnetic field $(eB)$ on the low-energy dynamics of quantum chromodynamics (QCD) is being extensively explored using different techniques: lattice QCD (LQCD)~\cite{DElia:2010abb, Bali:2011qj, Bali:2012zg}, MIT bag model~\cite{Chakrabarty:1996te, Fraga:2012fs}, functional renormalization group~\cite{Andersen:2013swa, Kamikado:2013pya}, chiral perturbation theory~\cite{Shushpanov:1997sf, Werbos:2007ym}, Nambu\textendash Jona-Lasinio (NJL) model~\cite{Ebert:1999ht, Inagaki:2003yi, Menezes:2008qt,Boomsma:2009yk,Ferrer:2013noa}, quark-meson model~\cite{Fraga:2008qn, Ruggieri:2013cya}, Polyakov loop extended models~\cite{Gatto:2010qs, Mizher:2010zb} etc. The major motivation for such studies comes from heavy-ion collision (HIC) experiments, where a magnetic field of strength $\sim 0.02-0.3\, {\rm GeV}^2$ can be generated depending on the collision energy~\cite{Skokov:2009qp}. Other relevant physical scenarios could include magnetars~\cite{Ng:2010zh, An:2014cwa, Kaspi:2017fwg} or the early epoch after the big bang~\cite{Widrow:2002ud}. 

Magnetic catalysis (MC) and inverse magnetic catalysis (IMC) are two of the major phenomena that we observe in such a magnetized QCD medium~\cite{Bali:2011qj, Bali:2012zg, Pagura:2016pwr, GomezDumm:2017iex, Ali:2020jsy}. MC is defined as the increasing magnitude of the chiral condensate, which is the order parameter associated with the spontaneous breaking of chiral symmetry, with increasing $eB$ and is typically seen at temperatures well below the crossover temperature $(T_{\text{CO}})$. On the other hand, the magnitude of the chiral condensate decreases with increasing $eB$ around the $T_{\text{CO}}$, and this is termed the IMC effect. It is to be noted that the IMC effect and a decrease in $T_{\text{CO}}$ with increasing $eB$ occasionally coexist. This is the reason that in much of the existing literature, the decrease in $T_{\text{CO}}$ with respect to the magnetic field is often referred to as the IMC effect.

However, this terminology is imprecise since one can devise scenarios where the chiral condensate around the $T_{\text{CO}}$ increases with increasing $eB$, but the $T_{\text{CO}}$ still shows the decreasing trend, showing that these two effects need not go together. This has been detailed in Ref.~\cite{DElia:2018xwo}. There, the authors explored the QCD phase diagram in the $T-eB$ plane for different pion masses. They observed that at pion masses heavier than the physical value, the IMC effect, the decreasing nature of the condensates with increasing $eB$, disappears, but the decreasing trend of $T_{\text{CO}}$ persists. In Ref.~\cite{Endrodi:2019zrl}, these findings were supported and further extended to find the maximum value of the pion mass beyond which the IMC effect disappears. All these discussions, along with other important updates of the field, have been recently reviewed in a recent review article~\cite{Endrodi:2024cqn}. The disappearance of the IMC effect can also occur in other circumstances, such as in a magnetized nonextensive QCD medium~\cite{Islam:2023zpl}. However, the phase diagram in the $T-eB$ plane shows non-monotonic behavior depending on the strength of the $q$-parameter, which defines the nonextensiveness of the system.

Two important points arise from the above discussion in Ref~\cite{DElia:2018xwo,Endrodi:2019zrl}—the importance of studying observables beyond the physical points to understand the structure of the underlying theory. The second point, which is associated with the first one, is whether some properties (such as the IMC in the present case) are intrinsic properties linked to the structure of the theory as we push beyond physical values.

In the quest for an analytical understanding of these observations, we use an effective QCD model that captures the global symmetries of QCD. Such models are simple but powerful, particularly for acquiring a qualitative understanding of the underlying theory. There exist many such models, such as the NJL model~\cite{Nambu:1961tp, Nambu:1961fr}, the linear sigma model~\cite{Gell-Mann:1960mvl}, etc. 

We will be using the NJL model in both its local and nonlocal avatars. Though one-to-one correspondence from these two methods (LQCD and effective models) is not feasible, we intend to use LQCD observation as a benchmark and test the models qualitatively. Such an exercise aims to lead to a better understanding of these effective models, exploring how they capture the essence of low-energy QCD and checking their validity beyond the physical point to capture the underlying structure.

In previous work, both the local~\cite{Farias:2014eca, Ferreira:2014kpa, Farias:2016gmy} and the nonlocal~\cite{Pagura:2016pwr, Ali:2020jsy} models were used to analyze the phase diagram of QCD at the physical pion mass in the presence of the magnetic field. To obtain IMC~\cite{Bali:2011qj, Bali:2012zg} near the crossover, the key ingredient is the weakening of the coupling with increasing energy. In the local models, it is incorporated via a $eB$ and $T$ dependent coupling constant. In the nonlocal NJL model, it comes naturally via an energy-dependent form factor whose strength reduces significantly at and above a characteristic scale $\Lambda$. However, the disappearance of the IMC effect beyond the physical point and the subsequent phase diagram in the $T-eB$ plane~\cite{DElia:2018xwo,Endrodi:2019zrl} provide a nontrivial challenge for such effective models.

Keeping that in mind, in the present paper, for the first time in an effective model framework, we study the effect of increasing the current quark mass (and hence the pion mass) on the phase diagram in the presence of the magnetic field. We know that the chiral condensate $(\langle\bar\psi\psi\rangle)$ is the observable (the order parameter) to explore the chiral dynamics of the system, used both in LQCD and effective model studies. We investigate this quantity to estimate the effects of the pion mass.

To realize the heavy pion mass in effective models, we vary the current quark mass while keeping the scale, $\Lambda$, and the coupling constant, $G$, fixed. This is motivated by the fact that the low-energy dynamics should not affect the scale of the theory, $\Lambda$. Hence, in effective models, changing only the current quark mass is a well-accepted way to achieve unphysical pion masses, whether to attain the chiral limit or to go beyond physical pion masses~\cite{Fukushima:2008wg, Nickel:2009wj}. This, of course, modifies the condensate and the pion decay constant accordingly, satisfying the relevant equations.

Our major observation is that with increasing pion mass, the IMC effect disappears, but the decreasing trend of $T_{\text{CO}}$ persists. This is qualitatively consistent with the LQCD results~\cite{DElia:2018xwo,Endrodi:2019zrl} and does not require the introduction of any new parameters. However, the value of the pion mass at which the IMC effect goes away is lower than that found in the LQCD studies~\cite{DElia:2018xwo,Endrodi:2019zrl}. In the nonlocal model, the IMC effect disappears around $m_\pi=215$ MeV. On the other hand, for the $2+1$-flavor local model, it disappears between $220$ and $340$ MeV of pion mass, except for $eB=0.2\, {\rm GeV}^2$ for which the IMC effect survives up to the highest pion mass value we tested. 

This observation is true for both the nonlocal and local models. Although, for the local model, the agreement depends on how the parameters of the model are fit at the physical point. For example, in the $2$-flavor local model, it is difficult to make any conclusive comment on the status of the IMC effect beyond the physical point, and the trend of the crossover temperature is non-smooth for higher $m_\pi$-values. On the other hand, in the $2+1$-flavor local model, the IMC effect for higher $m_\pi$-values shows irregular pattern for $eB$-dependence. In contrast, the IMC effect for beyond physical $m_\pi$-value shows a regular pattern in the nonlocal model. 

Inspired by the LQCD study~\cite{Endrodi:2019zrl}, we also perform an analysis to determine the pion mass value above which the IMC effect around the transition region exists no more. Being the most consistent model to capture the beyond physical point observations, we use the nonlocal framework for the purpose. We analyze for the whole range of $eB$ that we explored and find a pion mass value of $215$ MeV with a $1.4\%$ spread with a very small $eB$ dependence.

Finally, we try to understand the observations by looking at the characteristics of the effective coupling in these frameworks. In a nonlocal model, with its running of coupling, the IMC effect is captured automatically. On the other hand, in the local
models, one needs to introduce a parameterized running of the coupling constant to reproduce the IMC effect. Our findings suggest that the running of the coupling enables models to describe the effects of heavier current quark
masses without introducing additional parameters. We conclude that a smooth decrease of the effective coupling (with energy) via a form factor in the nonlocal model captures the physics more naturally.

The present paper is organized as follows. In the formalism section~\ref{sec:formalism}, we briefly discuss both the local and nonlocal NJL models. We start with the nonlocal version and then lead our discussion to the local ones. The essence of the parameter fitting is also discussed in detail in the respective subsections. We discuss our results in the section~\ref{sec:results} and compare them with LQCD results wherever appropriate. We also check the model's consistency in producing the results in subsection~\ref{ssec:cond_diff} and in subsection~\ref{ssec:coup_cons}, we try to understand them by exploring the running of the coupling constants in different formalisms. Finally, we conclude in section~\ref{sec:summary}.

%%%%%%%%%%%%%%%%%%%%%%%%%%%%%%%%%%%%%%%%%%%%%%%%%%%%%%%%%%%%%%%%%%%%%%%%
\section{Formalism}
\label{sec:formalism}
%%%%%%%%%%%%%%%%%%%%%%%%%%%%%%%%%%%%%%%%%%%%%%%%%%%%%%%%%%%%%%%%%%%%%%%%
As discussed, we will explore the effect of changing the current quark mass to obtain heavier than physical pion masses to test the effect on the Nambu\textendash Jona-Lasinio model in the presence of an external magnetic field. We investigate both the local and nonlocal versions. We will start our discussion with the nonlocal version~\cite{Bowler:1994ir, Schmidt:1994di, Plant:1997jr, GomezDumm:2001fz}. It is straightforward to reduce it to the local one by choosing an appropriate form factor. 

\subsection{NJL model with nonlocal interaction}
\label{ssec:form_nonlocal}
The Lagrangian for a $2$-flavor nonlocal NJL model is given by~\cite{Ali:2020jsy}
\begin{equation}
{\cal L}_{\text{NJL}}={\cal L}_0+{\cal L}_{\rm sym}+{\cal L}_{\rm det},
\label{eq:NJL}
\end{equation}
where different pieces are given as
\begin{equation}
\begin{split}
{\cal L}_0&=\bar{\psi}\left( i\slashed{\partial}-m\right)\psi,\\
{\cal L}_{\rm sym}&=G_{1}\left\{j_a(x)j_a(x)+\tilde{j}_a(x)\tilde{j}_a(x)\right\}\,\,{\rm and}\\
{\cal L}_{\rm det}&=G_{2}\left\{j_a(x)j_a(x)-\tilde{j}_a(x)\tilde{j}_a(x)\right\},
\end{split}
\label{eq:lag_each_term}
\end{equation}
where $\psi$ represents the light quark doublet containing the {\it u} and {\it d} quarks with equal current quark masses, $m$. ${\cal L}_0$ is the kinetic term, whereas ${\cal L}_{\text{sym}}$ and ${\cal L}_{\text{det}}$ are the interactions. ${\cal L}_{\text{det}}$ is the term responsible for explicitly breaking the $U(1)_A$ symmetry and incorporating the QCD axial anomaly in the effective model treatment. It is known as the 't Hooft determinant term, and its exclusion makes the total Lagrangian symmetric under the $SU(2)_V\times SU(2)_A\times U(1)_V\times U(1)_A$ transformation. The previous statements become evident as we make the currents explicit
\begin{eqnarray}
j_{a}(x)/\tilde{j}_a(x)=\int d^4z\ {\cal H}(z)\bar{\psi}\left(x+\frac{z}{2}\right)\Gamma_{a}/\tilde{\Gamma}_a\psi(x-\frac{z}{2}),
\label{eq:current}
\end{eqnarray}
where, $\Gamma=(\Gamma_0,\vec{\Gamma})=(\mathbb{I},i\gamma_5\vec{\tau})$, $\tilde{\Gamma}=(\tilde{\Gamma}_0,\vec{\tilde{\Gamma}})=(i\gamma_5,\vec{\tau})$, with $\vec{\tau}=(\tau^1,\tau^2,\tau^3)$ being the Pauli matrices and ${\cal H}(z)$, the nonlocal form factor in position space. It is easy to notice that with ${\cal H}(z)=\delta(z)$, one arrives at the currents for a local NJL model, which we discuss in detail in the section dedicated to the local versions.

In an isospin and parity symmetric scenario, the model properties are only dependent on $G_{1}+G_{2}$, as will be discussed below. Hence, the coupling constants $G_{1}$ and $G_{2}$ need not be equal (which is assumed in the standard NJL model, with only scalar and psuedoscalar-isovector interactions). They can be parameterized via a parameter, $c$, measuring the strength of the 't Hooft interaction term~\cite{Frank:2003ve,Ali:2020jsy,Ali:2021zsh}. In fact, in an isospin symmetry-breaking scenario, the inequality between the two coupling constants makes more sense, as found in Ref.~\cite{Ali:2020jsy} with $c=0.149\pm0.103$. However, in the present analysis, we are confined to discussing only isospin symmetric observables like condensate or mean-field averages, which are not very sensitive to $c$ as shown in Ref.~\cite{Ali:2020jsy}. Thus, we stick to the usual choice of $c=1/2$. Hence the coupling constants $G_{1}=G_{2}=G_{0}/2$, removing the interaction through $\tilde{j}_{a}(x)$ currents from the Lagrangian.

The free energy is the most essential thing to obtain in such an effective model framework. But as it stands, calculating the partition function using Eq.~\eqref{eq:NJL} is quite involved. One needs to linearize the theory through known techniques such as Hubbard-Stratonovich (HS) transformation~\cite{Hubbard:1959ub}. HS transformation introduces two auxiliary fields associated with the two different types of interactions that we have.
\begin{equation}
\begin{split}
\sigma(x)&=-\, \frac{G_{0}}{2}\int d^4z\ {\cal H}(z)\bar{\psi}\left(x+\frac{z}{2}\right)\Gamma_{0}\,\psi\left(x-\frac{z}{2}\right),\\
\vec{\pi}(x)&=-\, \frac{G_{0}}{2}\int d^4z\ {\cal H}(z)\bar{\psi}\left(x+\frac{z}{2}\right)\vec{\Gamma}\,\psi\left(x-\frac{z}{2}\right).
\end{split}
~\label{eq:sigma}
\end{equation}
Here, we have used the notation $\sigma$ and $\pi$ to represent the scalar-isoscalar and pseudoscalar-isovector auxiliary fields, respectively. In the mean-field approximation, these fields can acquire nonzero expectation values (while respecting the symmetries of the theory) depending on the strength of the interaction coupling. As we have started with an isospin and parity symmetric Lagrangian, the mean-field values of the auxiliary fields associated with Lorentz pseudoscalar and isovector channels must be zero. In the absence of any isospin breaking, the only relevant operator is in the scalar-isoscalar channel, which we denote as $\sigma(x)$. Expanding the auxiliary fields $\sigma(x)$ and $\vec{\pi}(x)$ around their mean-field values
\begin{equation}
    \begin{split}
    \sigma(x)&=\bar{\sigma}+\delta\sigma(x)\\
    \vec{\pi}(x)&=\delta\vec{\pi}(x)
    \end{split}
    \label{eq:fluctuation}
\end{equation}
and assuming $\bar{\sigma}$ to be space-time independent, one can obtain the mean-field Lagrangian, which is bilinear in quark fields. With this quadratic Lagrangian in quark fields, it is straightforward to integrate out the fermionic degrees of freedom, and one obtains the free energy per unit volume as
\begin{align}
\Omega= -2N_fN_c\int \frac{d^4q}{(2\pi)^4}\ln\left[q^2+M^2(q)\right]
+\frac{\bar{\sigma}^2}{2G_0}.
~\label{eq:free_ene_B0}
\end{align}
The momentum-dependent constituent quark mass $M(q)$ is given by
\begin{equation}
M_{u}(q)=M_{d}(q)=M(q)=m+h(q)\bar{\sigma}\;,
~\label{eq:eff_mass_B0}
\end{equation}
where $h(q)$ is the nonlocal form factor in the momentum space. Before going into the details of it, let us talk about the free energy given in Eq.~\eqref{eq:free_ene_B0}. The integral in the first term is divergent if one considers all possible momentum modes of quarks. As here we explore the low energy dynamics of a system of quarks, it is a standard procedure to suppress the quark degrees of freedom with momentum larger than a specific scale. From chiral perturbation theory, this scale is found to be of the order of $4\pi F_{\pi}\sim 1$ GeV~\cite{Manohar:1983md}. Now, there are different ways of incorporating this scale, which then render the free energy finite. The most used procedure is to introduce this scale as the cutoff in the $3$-momentum of the quark degrees of freedom~\cite{Klevansky:1992qe, Hatsuda:1994pi}. There exist a couple of other known procedures: the scale can be implemented as the covariant four-momentum cutoff or via the mass of a fictitious heavy particle in Pauli–Villars regularization or through an infrared red cut-off in the proper time regularization. A detailed discussion on the effect of regularization in the NJL model can be found in Ref.~\cite{Kohyama:2015hix}.

In the nonlocal model, we have the momentum-dependent form factor $h(q)$, which contains information about the scale of the theory. Here, we will use the Gaussian form factor, $h(q)=e^{-q^2/\Lambda^2}$, following the procedures in Refs.~\cite{GomezDumm:2006vz,Ali:2020jsy}, where $\Lambda$ represents the scale of the theory. One can have a different form factor, other than the Gaussian, as considered in Ref.~\cite{GomezDumm:2006vz}. Again, to put it into context, for local interaction, $h(q)=1$, and one needs an explicit cutoff to ignore/suppress the large momentum modes to make the free energy finite. An important point to note is that the nonlocal form factor does not make the free energy finite (as we are not allowed to put a cutoff in the momentum integration). The standard technique is to subtract an identical term from the equation with the effective quark mass, $M$, being replaced by its current counterpart, $m$~\cite{Hell:2008cc}. 

To study the chiral dynamics and vacuum properties, one uses the principle of least action to obtain the ground state of the system. In other words, the free energy is minimized with respect to the meanfield $(\bar{\sigma})$, i.e., by solving $\frac{\partial \Omega}{\partial\bar{\sigma}}=0$. It is straightforward to use Eq.~\eqref{eq:free_ene_B0} to obtain,
\begin{eqnarray}
\bar{\sigma} = 8 N_c \ G_0 \int \frac{d^4 q}{(2 \pi)^4}\  h(q) \ \frac{M(q)} {q^2 + M^2(q)}\;.
\label{eq:gapeq_B0}
\end{eqnarray}
On the other hand, to obtain the chiral condensate, one can use the Feynman-Hellmann theorem~\cite{Hell:2008cc} by differentiating the free energy with respect to the current quark mass, $m$
\begin{eqnarray}
\langle\bar{\psi}_f(x)\psi_f(x)\rangle=\frac{\partial\Omega}{\partial m} &=& - \, 4 N_c \int \frac{d^4 q}{(2 \pi)^4}\left[\frac{M(q)} {q^2 + M^2(q)}-\frac{m} {q^2 + m^2}\right].
\label{eq:cond_B0}
\end{eqnarray}
There are a few important things to note here. The minimization condition [Eq.~\eqref{eq:gapeq_B0}], known as the gap equation, is convergent with appropriate nonlocal form factors. On the other hand, the condensate, given by Eq.~\eqref{eq:cond_B0}, is divergent without the subtraction of the second term. The subtraction process might appear ad hoc, but Eq.~\eqref{eq:cond_B0} can be obtained using the Feynman-Hellmann theorem on the regularized free energy. For the local scenario, one simply uses a cutoff in the integration for both cases, which will be fixed to produce physical observables. The second part of the Eq~\eqref{eq:cond_B0} does not appear as the subtraction is absent in the first place. We will recall this subtraction technique once again in the next section, where we learn that the subtraction is to be done carefully in the presence of a magnetic field.

Another important thing to note is that the direct proportionality between the meanfield and the condensate no longer exists in a nonlocal NJL model, contrary to its local counterpart. This is obvious from a comparison between Eqs.~\eqref{eq:gapeq_B0} and~\eqref{eq:cond_B0}. This introduces some subtlety in the computation procedure and will be highlighted wherever deemed necessary. So far, the discussion is for the vacuum. To serve our purpose, we need to include both the temperature and the magnetic field, which we will do in the next to next section. Before that, we provide the details of the parameter fitting, which plays a central role in our discussion.

\subsection{Parameter fitting}
\label{ssec:para_fit}
To fit the model parameters, we use condensate $(\langle\bar\psi\psi\rangle)$, pion mass $(m_\pi)$ and pion decay constant $(F_\pi)$. The fitted parameters are the current quark mass $(m)$, the coupling constant $(G_{0})$, and $\Lambda$. It is to be noted that $\Lambda$ is not exactly the cut-off we encounter in the usual NJL model. Rather, $\Lambda$ works as a scale that characterizes the range of the nonlocal interaction, beyond which the coupling constant becomes smaller and smaller with increasing momentum.

To study the mesonic properties within the NJL framework, one can expand the mean-field action in powers of the fluctuations [see Eq.~\eqref{eq:fluctuation}] associated with the auxiliary fields around their meanfield values. The coefficient of the quadratic term in these fluctuations can be identified as the inverse of the propagator of the respective modes. The inverse propagator for the $\sigma$ and $\pi^{0}$ modes are given by~\cite{GomezDumm:2006vz}
\begin{eqnarray}
{\cal G}^{\pm}(p^2) = \frac{1}{G_0} - \, 8 \,N_c \int \frac{d^4 q}{(2 \pi)^4}\ h^2(q) \frac{  \left[ (q^+ \cdot q^-) \mp M(q^+) M(q^-)\right]}{\left[ (q^+)^2 + M^2(q^+) \right] \left[ (q^-)^2 + M^2(q^-)\right]},
\label{eq:pion_prop}
\end{eqnarray}
where $\pm$ signs in ${\cal G}^{\pm}$ stands for sigma mode and pion mode, respectively; with $q^{\pm}=q\pm p/2$. We are interested in the pion mode, and the pion mass is obtained by equating
\begin{eqnarray}
{\cal G}^{-}(-m_\pi^2) = 0 \ .
\label{eq:pion_mass}
\end{eqnarray}
The pion decay constant ($F_{\pi}$) can be extracted from the one pion to the vacuum matrix element of the axial vector current ($J_{\mu5}^{a}$) given below,
\begin{equation}
    \langle0|J_{\mu5}^{a}(0)|\pi^{b}(p)\rangle=i\delta^{ab}F_{\pi}p_{\mu}.
    \label{eq:fpi_identity}
\end{equation}
We follow the steps in Ref.~\cite{GomezDumm:2006vz} and use the following expression for $F_\pi$
\begin{equation}
m_\pi^2 \; F_\pi  = m \; Z_\pi^{1/2}\; J(-m_\pi^2) \ ,
\label{eq:exp_fpi}
\end{equation}
with $J(p^2)$ is given by
\begin{eqnarray}
J(p^2) = \, 8 \,N_c \int \frac{d^4 q}{(2 \pi)^4}\ h(q) \frac{  \left[ (q^+ \cdot q^-) + M(q^+)
M(q^-)\right]}{\left[ (q^+)^2 + M^2(q^+) \right] \left[ (q^-)^2 + M^2(q^-)\right]}\,,
\label{eq:jp_square}
\end{eqnarray}
and $Z_\pi$ is connected to the $\pi\bar{\psi}\psi$ coupling constant and is given as
\begin{equation}
Z_\pi^{-1} = \frac{d {\cal G}^-(p) }{dp^2}
\bigg|_{p^2=-m_\pi^2} \ .
\label{eq:zpi}
\end{equation}
We use Eqs.~\eqref{eq:cond_B0},~\eqref{eq:pion_mass} and~\eqref{eq:exp_fpi} to fit the model parameters $m$, $G_{0}$ and $\Lambda$ to reproduce the empirical values of the pion mass, pion decay constant, and the chiral condensate. In this study, we change the current quark mass, $m$, to get to the values of unphysical pion masses. While doing this, we keep $G_{0}$ and $\Lambda$ fixed (to their corresponding values at the physical point), which is a widely used procedure to go to unphysical pion masses (e.g., chiral limit, $m=0$) in effective model framework. Changing $m$ will change the other observables, though the condensate and pion decay constant are less sensitive than the pion mass. A similar procedure has been implemented for the local framework as well. The parameter fitting will be discussed in detail in the result section.

\subsection{Introduction of the temperature and the magnetic field}
\label{ssec:form_T_eB}
This section briefly discusses the procedure to incorporate the effect of the temperature and magnetic field. Since the Refs.~\cite{Pagura:2016pwr, Ali:2020jsy} contains all the necessary details about the derivations of the free energy, gap equations, condensates, etc., within a nonlocal NJL model in the presence of temperature and magnetic field. We will keep only those directly associated with the results discussed in this article. The major difference between the Ref.~\cite{Ali:2020jsy} and the present study is that here, we have only one mean-field, $\bar{\sigma}$, because of our choice to work with $c=1/2$. This simplifies some of the computational complexity.

To calculate the chiral condensate in the presence of the magnetic field, once again, we use the Feynman-Hellmann theorem, which is similar to the case of zero magnetic fields. The differentiation of the free energy with respect to $m$ leads to
\begin{eqnarray}
\langle\bar\psi_f\psi_f\rangle & = & - N_c \frac{2|q_f B|}{2 \pi} \int \frac{d^2q_\parallel}{(2\pi)^2} \ \Bigg\{ \frac{{M^{s_{\! f},f}_{q_\parallel,0}\,}}{q_\parallel^2 + \left({M^{s_{\! f},f}_{q_\parallel,0}\,}\right)^2}+ \nonumber \\
& & \sum_{k=1}^\infty \frac{ \left( 2 k |q_f B| + q_\parallel^2 +
	M^{-1,f}_{q_\parallel,k} M^{+1,f}_{q_\parallel,k}\right)\left(
	M^{+1,f}_{q_\parallel,k} + M^{-1,f}_{q_\parallel,k} \right)}{ \left( 2 k |q_f B| + q_\parallel^2 +
	M^{-1,f}_{q_\parallel,k} M^{+1,f}_{q_\parallel,k}\right)^2 \! \! + q_\parallel^2 \left(
	M^{+1,f}_{q_\parallel,k} - M^{-1,f}_{q_\parallel,k} \right)^2} \Bigg\},
\label{eq:cond_B}
\end{eqnarray}
where $f$ stands for individual flavor $u$ and $d$ \footnote{The same equation in Ref.~\cite{Ali:2020jsy}, i.e., Eq. 37, has a typo. The factor ${\rm ln}\,x_f$ should be outside the parenthesis and multiplied with the term $(1-\frac{1}{2x_f})$.}. The constituent mass $(M^{\lambda,f}_{q_\parallel,k})$ in the presence of a magnetic field with a Gaussian form factor is 
\begin{equation}
M^{\lambda,f}_{q_\parallel,k} = m + \bar{\sigma} \ \frac{ \left(1- |q_f B|/\Lambda^2\right)^{k+\frac{\lambda s_{\! f}-1}{2}}} { \left(1+ |q_f B|/\Lambda^2\right)^{k+\frac{\lambda s_{\! f}+1}{2}}} \;\exp\!\big(-{q_{\parallel}}^{2}/\Lambda^2\big)\,,
\label{eq:eff_mass_B}
\end{equation} 
with $q_{\parallel}=(q_3,q_4)$, $k$ is the Landau level index, $\lambda=\pm1$ is the spin and $s_f={\rm sign}(q_f)$. To introduce the temperature, we use the Matsubara formalism~\cite{Matsubara:1955ws, Mustafa:2022got}, which connects the Euclidean time component of the momentum to the temperature following the identity $q_{4}=(2n+1)\pi T$, for fermions. There is subtlety in dealing with nonzero temperature in a nonlocal framework. As opposed to its local cousin, we cannot perform the sum analytically, and the vacuum and the thermal contribution are entangled. Thus, in a nonlocal setup, the Matsubara sum needs to be performed numerically.

The expression is divergent at large momentum with nonzero quark mass. To remove the divergence, we use the same technique as we did for zero magnetic fields [Eq.~\eqref{eq:cond_B0}]. We subtract the same term, but without any interaction, i.e., $M$ being replaced by $m$. The magnetic field and the temperature also remain present. We call it the ``free" term.

At this point, there arises an important difference as compared to the vacuum case [Eq.~\eqref{eq:cond_B0}]. There, subtracting the part with $m$ was sufficient, as this ``free" term is always infinite. But here, in the presence of a temperature and a magnetic field, that is not the case. The ``free" term contains a finite contribution, which we denote as ``free,reg" and needs to be added to the final expression~\cite{Pagura:2016pwr, Ali:2020jsy}. Thus, the regularized expression for the condensate, in the presence of temperature and magnetic field, is
\begin{equation}
\langle\bar\psi_f\psi_f\rangle^{\text{reg}}_{B,T}=\langle\bar\psi_f\psi_f\rangle_{B,T}-\langle\bar\psi_f\psi_f\rangle^{\text{free}}_{B,T}+\langle\bar\psi_f\psi_f\rangle^{\text{free,reg}}_{B,T},
\label{eq:cond_B_reg}
\end{equation}
where $\langle\bar\psi_f\psi_f\rangle^{\text{free,reg}}_{B,T}$ is given as~\cite{Menezes:2008qt}
\begin{eqnarray}
\langle\bar\psi_f\psi_f\rangle^{\text{free,reg}}_{B,T}&=&-\frac{N_cm^3}{4\pi^2}\Bigg[\frac{\ln\Gamma(x_f)}{x_f}-\frac{\ln(2\pi)}{2x_f}+1-\left(1-\frac{1}{2x_f}\right)\ln x_f\Bigg]\nonumber\\
&&+\frac{N_c\left|q_fB\right|}{\pi}\sum_{k=0}^{\infty}\alpha_k\int\frac{dq}{2\pi}\frac{m}{E^f_k\left(1+\exp[E^f_k/T]\right)}\,;
\label{eq:cond_B_free_reg}
\end{eqnarray}
with $x_f=m^2/(2\left|q_fB\right|)$ and $\alpha_k=2-\delta_{k0}$, is the Landau level degeneracy factor. Here, we should remind ourselves again that the regularized form of the condensate in Eq.~\eqref{eq:cond_B_reg} can be directly obtained from a regularized expression of the free energy. In that case, the free energy needs to be regularized in the same fashion as it is described for its zero magnetic field counterpart [Eq.~\eqref{eq:free_ene_B0}]. The first term in the above equation is a pure magnetic field-dependent term, for which the summation over the Landau levels is performed~\footnote{There are two typos in the term ``free,reg'' (Eq. 37) in Ref~\cite{Ali:2020jsy}. An overall minus sign is missing, and also ``$\ln x_f$'' within the square brackets is misplaced.}. 

\subsection{NJL model with local interaction}
\label{ssec:form_local}
It is very straightforward to arrive at the local version of the NJL model from the nonlocal one. By choosing the nonlocal form factor to be a delta function ${\cal H}(z)=\delta(z)$, the interaction becomes point-like (local). Thus, to get to the $2$-flavor local NJL model, one needs to replace the form factor accordingly in Eq.~\eqref{eq:current}. In momentum space, this choice leads to $h(q)=1$, and one needs to put an explicit cutoff in the momentum integration. Important to note that the subtraction implemented to make the free energy finite in the nonlocal scenario is no longer required with the introduction of the explicit cutoff. 

To study the $2$-flavor local NJL model in a magnetic field, we follow the framework given in Ref.~\cite{Farias:2016gmy}, which can be easily implemented following the above arguments. Thus, we do not repeat them here. To obtain the IMC effect, they fitted the coupling constant. For a better understanding of the result, we rewrite here its form,
\begin{align}
 G_0(eB,T)=c(eB)\left(1-\frac{1}{1+e^{\beta(eB)(T_a(eB)-T)}}\right)+s(eB),
 \label{eq:coupling_2}
\end{align}
where the parameters $c$, $\beta$, $T_a$ and $s$ depend on only the external magnetic field. Their fitted values can be found in Ref.~\cite{Bandyopadhyay:2023lvk}\footnote{The values of the parameter $\beta$ in Table 1 in Ref.~\cite{Farias:2016gmy} are misquoted. The right numbers are given in Ref.~\cite{Bandyopadhyay:2023lvk}.}. In our analysis, we do not change Eq.~\eqref{eq:coupling_2} as we change the current quark mass.

The above argument also holds for a $3$-flavor model. The NJL model with $3$ flavors of quarks is a bit more complicated with six-quark interaction, but within the mean-field approximation, the free energy and other expressions are analogous to those of the $2$ flavors. 

The basic structure of the Lagrangian is the same as in Eq.~\eqref{eq:NJL} with some pertinent modifications. The Dirac spinor is now a triplet with $\psi^{\text{T}}=(u,d,s)$ and their masses make the mass matrix $\hat{m}=\text{diag}(m_{u},m_{d},m_{s})$. The symmetries of the the interaction terms ${\cal L}_{\text{sym}}$ and ${\cal L}_{\text{det}}$ are similar to $2$-flavor with $N_f=2$ being replaced by $N_f=3$. The explicit forms of these interactions are given below~\cite{Klevansky:1992qe,Hatsuda:1994pi}
\begin{equation}
\begin{split}
  {\cal L}_{\text{sym}}&=G_{1}\sum_{a=0}^{8}\left[\left(\bar{\psi} \lambda_a\psi\right)^2 + \left(\bar{\psi}\,i \gamma_5 \lambda_a\psi\right)^2\right]\; {\rm and} \\
  {\cal L}_{\text{det}}&=- G_{2}\left[\det\bar{\psi}_i (1-\gamma_5)\psi_j+\det\bar{\psi}_i (1+\gamma_5)\psi_j\right],
\end{split}
  \label{eq:NJL_Interaction}
\end{equation}
where the generators of $SU(3)$ symmetry are represented by the Gell-Mann matrices $\lambda$, and the determinant in ${\cal L}_{\text{det}}$ is taken in the flavor space. The coupling constants are denoted as $G_{1}$ and $G_{2}$ to signify the similar physics that the corresponding interaction terms represent in both $2$- and $3$-flavors. However, their explicit values, as mentioned below, will be different. Once we have the Lagrangian, we follow the same procedure as described earlier to derive the free energy. 

To introduce the magnetic field, we exactly follow the procedure of Ref.~\cite{Ferreira:2014kpa}. Our goal is to apply the framework developed there to test its validity beyond the physical point. Thus, we skip repeating the formalism here. However, to comprehend the result, we would require the form of the coupling constant, as we did for the $2$-flavor. It is fitted as 
\begin{align}
 G_{1}(\xi)=G_{1}^{0}\,\frac{1+a\xi^2+b\xi^3}{1+c\xi^2+d\xi^4},
 \label{eq:coupling_2plus1}
\end{align}
with $a = 0.0108805$, $b = -1.0133\times10^{-4}$, $c = 0.02228$, $d = 1.84558\times10^{-4}$; $\xi=eB/\Lambda_{\text{QCD}}^2$ with $\Lambda_{\text{QCD}}=0.3$ GeV, and $G_{1}^{0}=G_{1}(eB=0)=G_{1}$. Note that the external field dependence is not considered for the other coupling, $G_{2}$. The $eB$-dependence of the 't Hooft determinant term has been explored in Ref.~\cite{Moreira:2020wau}.

\subsection{Parameter fitting for a $2+1$ flavor NJL model}
The parameter fitting in a $3$-flavor NJL model is different and needs a separate discussion. It has a total of $5$ parameters that need to be fitted to produce the empirical values of the observables like meson masses, their decay constants, etc. The $5$ parameters are the light quark mass, $m$ (considering isospin symmetry), the strange quark mass, $m_{s}$, the two coupling constants, $G_{1}$ and $G_{2}$, and a three momentum cutoff $\Lambda$. The most used parameter sets, from Refs.~\cite{Hatsuda:1994pi, Rehberg:1995kh}, are obtained by fitting $m_{s}$, $G_{1}$, $G_{2}$, and $\Lambda$ to reproduce the experimental values of masses of pion, kaon, and $\eta'$ mesons and the pion decay constant while the light quark mass, $m$ is chosen to be $5.5$ MeV as obtained in chiral perturbation theory~\cite{Gasser:1982ap}.

The detailed calculation of the different meson propagators and decay constants are given in Refs.~\cite{Hatsuda:1994pi, Rehberg:1995kh}. Going to the unphysical region in a $2+1$ flavor NJL model is not as straightforward as for the $2$ flavor case. In other words, simply changing the light quark mass while keeping all other parameters fixed will not work due to the presence of kaon and $\eta$ mesons. For example, for a pion mass larger than the physical value, if we only change the light quark mass, the pions and kaons will become degenerate at $m=m_{s}$. In order to maintain the correct mass hierarchy between pions and kaons, it is necessary to adjust the strange quark mass as well. The procedure for that is to keep the ratio of light to strange quark mass fixed to the physical value while going to the unphysical region by changing quark masses~\cite{Fukushima:2008wg, DElia:2018xwo}. We further discuss it in the results section.

%%%%%%%%%%%%%%%%%%%%%%%%%%%%%%%%%%%%%%%%%%%%%%%%%%%%%%%%%%%%%%%%%%%%%%%%
\section{Results}
\label{sec:results}
%%%%%%%%%%%%%%%%%%%%%%%%%%%%%%%%%%%%%%%%%%%%%%%%%%%%%%%%%%%%%%%%%%%%%%%%
We divide our results into two major sub-sections: nonlocal (Sec.~\ref{ssec:nonlocal}) and local (Sec.~\ref{ssec:loc_mod}). We show the result for the nonlocal version first and then switch to the local one. We start with reviewing the parameters of the two models and how they are fitted. We then describe our findings on individual models. 

In a later sub-section (Sec.~\ref{ssec:cond_diff}), we investigate the correlation between the IMC effect around the phase transition temperature and the decreasing $T_{\text{CO}}$ by looking at the condensate-average differences. To avoid confusion, we note that this is different from the condensate difference, in which the difference between the $u$ and $d$-quark condensate is calculated. The condensate-average difference is defined as the average of the $u$ and $d$ quark condensates from which the value of the condensate average for $eB=0$ is subtracted. This helps to recognize the IMC effect more clearly. In Sec.~\ref{ssec:coup_cons}, we try to understand all the results by dissecting the model ingredients\textemdash\, mainly the running of the coupling constants.

\subsection{Nonlocal version}
\label{ssec:nonlocal}
We begin by describing the parameter fitting. For the nonlocal model, the physical point corresponds to the LH parameter set of Ref.~\cite{Ali:2020jsy} with $m_{\pi}=135$ MeV, $F_{\pi}=92.9$ MeV and $\langle\bar{\psi}_{i}\psi_{i}\rangle^{1/3}=221.1$ MeV. The model parameters $m$, $G_{0}$ and $\Lambda$ are fitted to obtain these obervables using Eqs.~\eqref{eq:cond_B0},~\eqref{eq:pion_mass} and~\eqref{eq:exp_fpi}. The details of the parameters can be found in Ref.~\cite{Ali:2020jsy}. 

To go to unphysical pion masses, we follow the most used procedure for effective models where one increases the current quark masses while keeping other parameters such as $G_{0}$ and $\Lambda$ fixed~\cite{Fukushima:2008wg, Nickel:2009wj}. The argument is that $\Lambda$, being the scale of the theory, should not be affected by the low energy dynamics. On the other hand, $G_0$, being the coupling constant, should not be affected by the varying current quark masses with a fixed $\Lambda$. Following these arguments, we solve Eqs.~\eqref{eq:gapeq_B0} and~\eqref{eq:pion_mass} consistently to obtain the desired pion masses. The unphysical pion masses and the corresponding current quark masses are mentioned in Table~\ref{tab:parameters}. They are obtained with $\Lambda$ and $G_{0}$ kept fixed to their values at physical point: $\Lambda=605.05$ MeV and $G_{0}=29.38/\Lambda^{2}$. The major motivation behind these chosen values comes from the LQCD study~\cite{DElia:2018xwo}. However, in LQCD, the analysis reaches pions above $600$ MeV, which is not feasible in the models considered here. \\
%%%%%%%%%%%%%%%%%%%%%%%%%%%%%%%%%%%%%%%%%%%%%%%%%%%%%%%%%%%%%%%%%%%%%%%%
\begin{table}[h]
    \centering
    \begin{tabular}{|c|cccccc|}
        \hline
        $m_\pi\, ({\rm MeV})$& & $135$  & $220$ & $340$ & $440$ & \\ \hline
        & Nonlocal $2f$  : & $6.9$ & $18.3$  & $42.9$  & $70.8$ & \\ 
        $m$ ({\rm MeV}) & Local $2f$: & $5.5$ & $13.2$  & $32.2$  & $54.2$ & \\
        & Local $2+1\,f$: & $5.5$ & $14.5$  & $34.1$  & $56.1$ &\\\hline 
        \end{tabular}
    \caption{Fitted $m$-values for different pion masses for all three models explored here.}
    \label{tab:parameters}
\end{table}

%%%%%%%%%%%%%%%%%%%%%%%%%%%%%%%%%%%%%%%%%%%%%%%%%%%%%%%%%%%%%%%%%%%%%%%%

To keep the present discussion in perspective, we would like to mention that with the parameter set for the physical point, one gets a reasonable match for the condensate and scaled crossover temperature as a function of $eB$ while comparing with LQCD results~\cite{Ali:2020jsy}. Here, we are focused on the IMC effect and the behavior of $T_{\text{CO}}$. Thus, calculating the condensate average will be enough for us. We also seek a qualitative understanding of our result and do not aim for a quantitative matching. Thus, we calculate $\langle\bar{\psi}_{i}\psi_{i}\rangle$ without any further scaling or subtraction, which is required to match with the LQCD result~\cite{Bali:2012zg, Ali:2020jsy} and present the average of the $u$ and $d$-quark condensates, denoted as $\langle\bar\psi\psi\rangle_{\rm Ave}$.

\begin{figure}[h!]
  \includegraphics[scale=0.8]{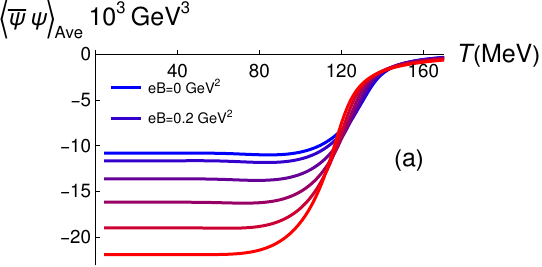}
  \includegraphics[scale=0.8]{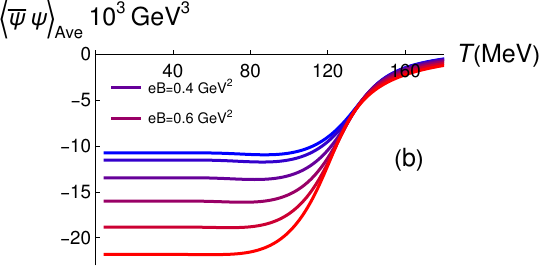}\\
  \includegraphics[scale=0.8]{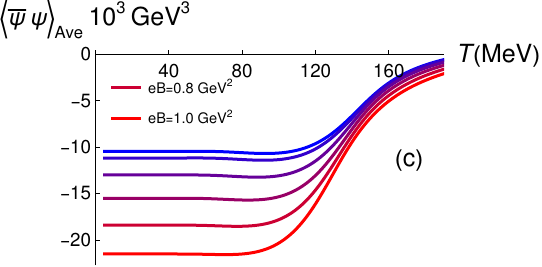}
  \includegraphics[scale=0.8]{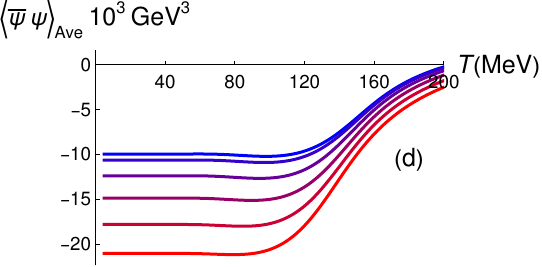}
  \caption{Condensate average as a function of temperature for different values of $eB$, with increasing strength from blue to red as denoted by the legends. The sub-figures (a), (b), (c), and (d) correspond to $m_\pi=135$, $220$, $340$, and $440$ MeV, respectively. The condensate average values are scaled with a factor of $10^{3}$.}
  \label{fig:cond_ave}
\end{figure}

Let us start discussing the result by looking at Fig.~\ref{fig:cond_ave}, where we present $\langle\bar\psi\psi\rangle_{\rm Ave}$ as a function of temperature for different values of the pion mass in the presence of a magnetic field. In the figure, the changing color from blue to red stands for the increasing strength of the magnetic field, with blue and red representing $0$ and $1$ ${\rm GeV}^2$ $eB$, respectively, in steps of $0.2\ \text{GeV}^{2}$. We observe that the effect of IMC starts to decrease as we increase the mass of the pion, which qualitatively resembles the observation in LQCD~\cite{DElia:2018xwo}. The IMC effect disappears for larger values of $m_\pi$. It happens at a much lower value in the present model as compared to the LQCD. At this point, one must look into the behavior of $T_{\text{CO}}$ as a function of $eB$ for different values of $m_\pi$.

\begin{figure}[h!]
  \includegraphics[scale=0.74]{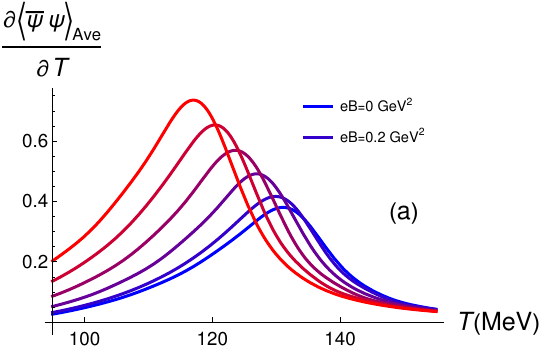}
  \includegraphics[scale=0.74]{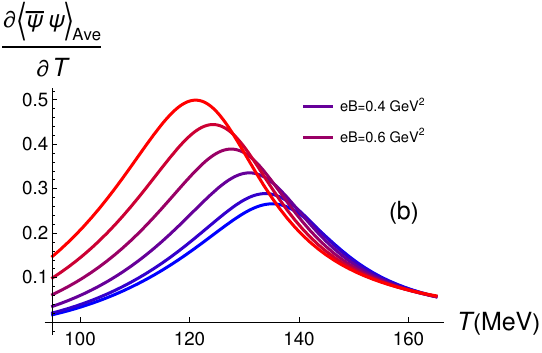}\\
  \includegraphics[scale=0.74]{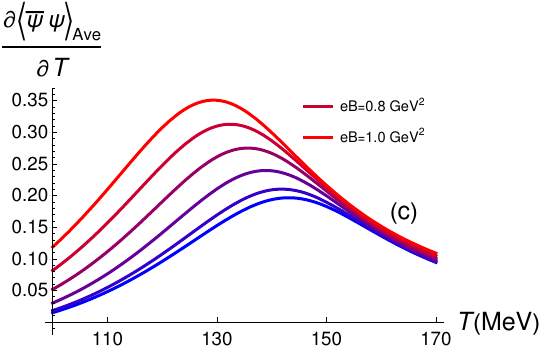}
  \includegraphics[scale=0.74]{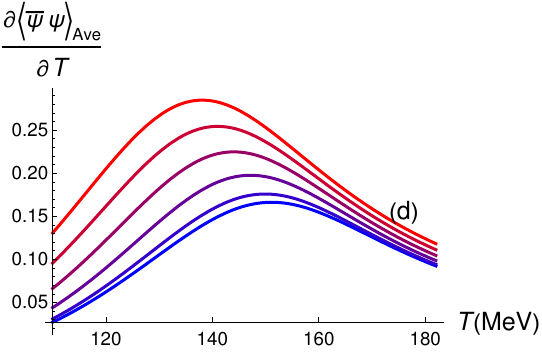}
  \caption{Temperature gradient of the condensate average as a function of temperature for different values of $eB$ with increasing strength from blue to red as denoted by the legends. The sub-figures (a), (b), (c), and (d) correspond to $m_\pi=135$, $220$, $340$, and $440$ MeV, respectively.}
  \label{fig:cond_grad}
\end{figure}

This is what Fig.~\ref{fig:cond_grad} displays. We follow the known method of calculating the pseudo-critical temperature for the restoration of the chiral symmetry by finding the inflection point from the temperature gradient of the condensate average. We have plotted the temperature gradient of the condensate average with increasing $m_\pi$ from left to right. The plots are shown at least once to give readers an understanding of the overall behaviors of such a quantity. It is obvious, in all the cases, that the peak of the gradient (signals the inflection point) shifts to the left with increasing strength of the magnetic field. This is what the authors of Ref~\cite{DElia:2018xwo} have observed in an LQCD simulation.

It is interesting to observe that simple effective models like NJL, constructed through symmetry arguments, are capable of capturing the major essence of an LQCD calculation. In this regard, the nonlocal version, introduced to capture the essence of the running of the strong coupling, is efficient. Previously, it was used without any further tweaking like its local counterpart~\cite{Farias:2014eca, Providencia:2014txa, Farias:2016gmy} to reproduce the IMC effect. Here, we extend it to test successfully whether it can be trusted in the region away from the physical point. Such an extension validates the most important motivations behind considering a nonlocal version of the NJL model~\cite{GomezDumm:2006vz, Hell:2008cc, GomezDumm:2017iex} showing its usefulness.

\begin{figure}[h!]
  \includegraphics[scale=0.39]{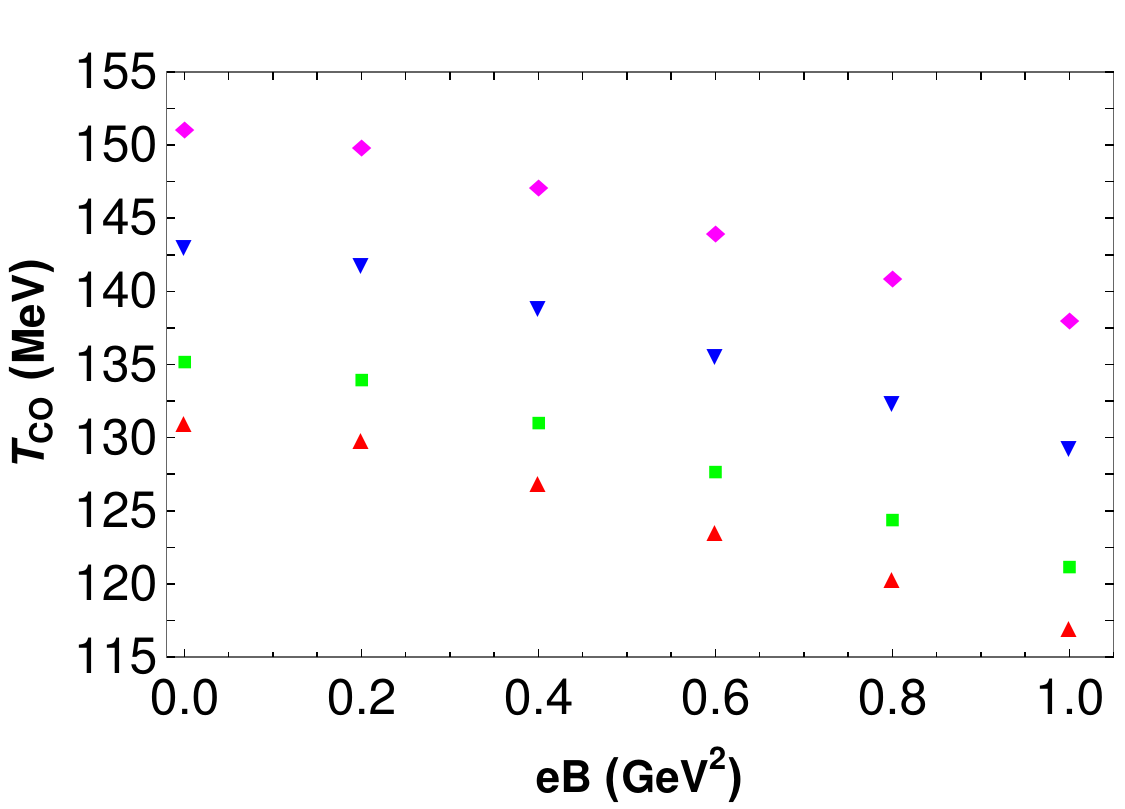}
  \includegraphics[scale=0.39]{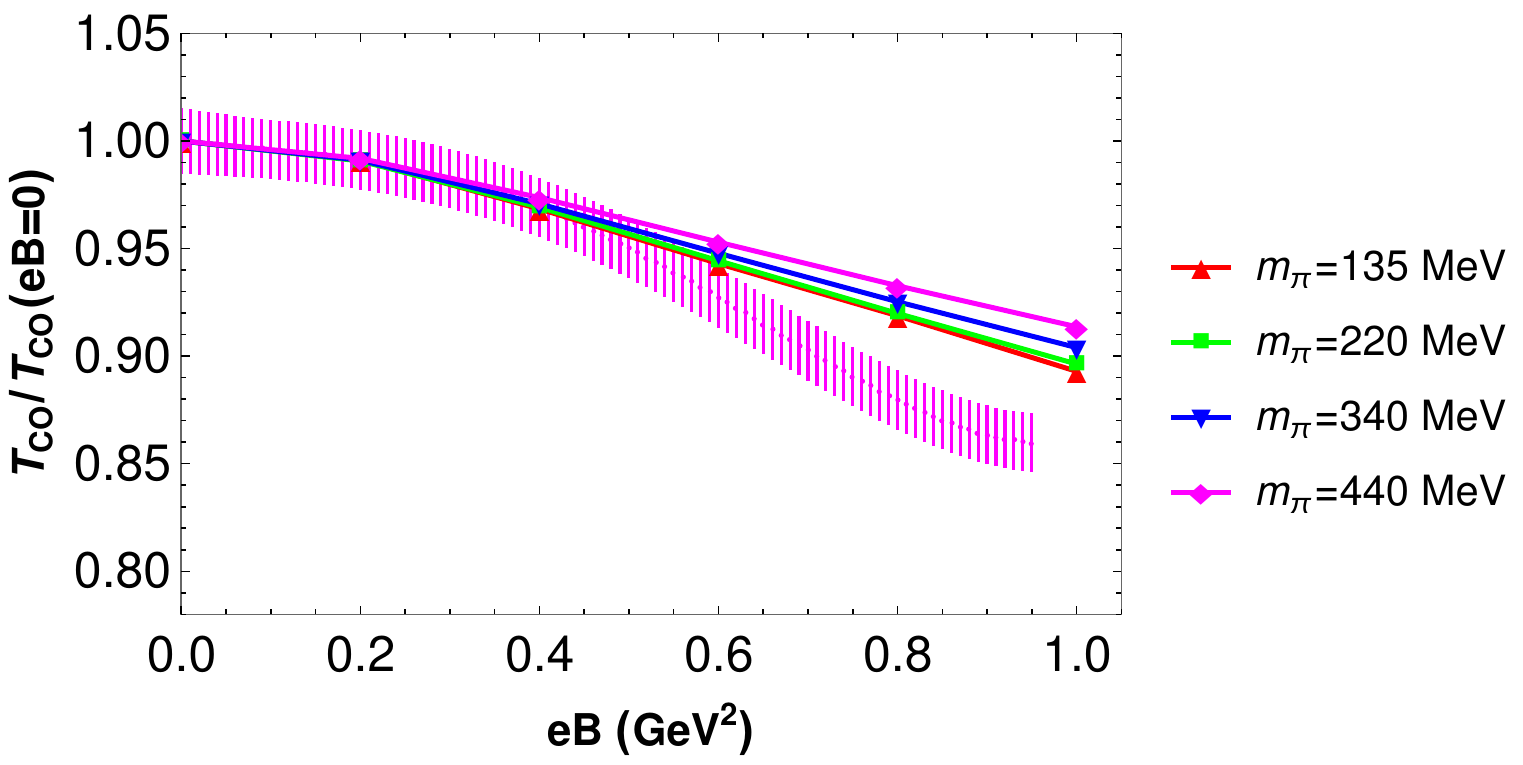}
  \caption{Left: Crossover temperatures as a function of $eB$ for different $m_\pi$. Right: The same plot with crossover temperatures scaled with respective zero $eB$-values and compared with LQCD result~\cite{Bali:2012zg} for physical pions, shown with the magenta band. The lines are solely intended to guide the eyes.}
  \label{fig:TcVseBdiffMpi}
\end{figure}

Let us draw the phase diagram for better visualization of the behavior of $T_{\text{CO}}$ for different pion masses. We have presented the crossover temperature as a function $eB$ for different values of the pion mass in Fig.~\ref{fig:TcVseBdiffMpi}. The left panel of the figure shows the values of the crossover temperature as a function of the magnetic field for different values of pion masses. With increasing pion mass, the values of the $T_{\text{CO}}$ increase, but with increasing strength of $eB$, it always has a decreasing trend. This was already evident from different panels of Fig.~\ref{fig:cond_grad}. However, here in Fig.~\ref{fig:TcVseBdiffMpi}, we can have a quantitative estimation along with a relative impact of different pion masses. This is similar to Fig. 11 in Ref.~\cite{DElia:2018xwo}.

To compare with the LQCD findings, we have presented the scaled crossover temperature as a function of the magnetic field for different values of pion masses in the right panel of Fig.~\ref{fig:TcVseBdiffMpi}. The LQCD data has been taken from Ref.~\cite{Bali:2011qj}. The result for the physical $m_\pi$ is known from the Ref.~\cite{Ali:2020jsy}, which we calculated here as well. With increasing $m_\pi$, the decreasing trend of the scaled transition temperature slightly goes up, especially for higher values of $eB$. This is also qualitatively similar to Fig. 12 in the Ref.~\cite{DElia:2018xwo}. 

\subsection{Local versions}
\label{ssec:loc_mod}
As mentioned earlier, there are a few ways to obtain the inverse magnetic catalysis effect in the presence of a magnetic field within the NJL framework in its local version. We focus on two local NJL model frameworks: one where the interaction strength is a function of both the magnetic field and temperature~\cite{Bandyopadhyay:2023lvk} and another where it is dependent only on the magnetic field~\cite{Ferreira:2014kpa}. However, both approaches rely on the main idea that the interaction strength should be energy-dependent.

\subsubsection{$T$ and $eB$ dependent coupling constant}
\label{sssec:coup_cons_T_eB}
One of the very first references to reproduce the IMC effect in a $2$-flavor NJL model was reported in Ref.~\cite{Farias:2014eca}. The running of coupling as a function of the external field and the temperature through some fitted parameters enables the model to capture such a phenomenon. The authors further improved on their fitting in their next investigation in Ref.~\cite{Farias:2016gmy}. The condensate averages determined by LQCD study~\cite{Bali:2012zg} are used as the observables to fit. This fitting~\footnote{There is a typo in Eq. (16) of Ref.~\cite{Farias:2016gmy}: $\psi^{-2}(x_f)$ should lie outside the square brackets, i.e., $x_f/2$ should not be multiplied with it.} has been extensively used to calculate other quantities in the presence of a magnetic field, for example, transport coefficients~\cite{Bandyopadhyay:2023lvk}.

Thus, we use the method of Ref.~\cite{Farias:2016gmy} for our purpose to go beyond the physical point and check its validity. As before, we begin with the parameter fitting that is required to attain higher pion masses. The method is similar to the $2$-flavor nonlocal version. We keep the coupling constant and the cut-off fixed and obtain higher pion masses with higher values of current quark masses.

Obviously, we take the parameter set for the physical point from Ref.~\cite{Farias:2016gmy} as the fittings are done for that set. Then, to go to the larger pion mass, we use expressions for the relevant observables such as $m_\pi$, $F_\pi$ and $\langle\bar\psi\psi\rangle$ in sync with the formalism used there. However, it is to be mentioned that the parameter values considered at the physical point are not exactly obtained by solving the relevant expressions. Rather, the values of $m_\pi$, $F_\pi$ and $m$ are chosen to match with the condensate value found by the LQCD study~\cite{Bali:2012zg} through the Gell-Mann-Oakes-Renner (GOR) relation. Then they consider a cut-off value, $\Lambda=650$ MeV, as a standard value for the model and fit the lattice-defined condensate average to calculate $G_{0}$ at zero $T$ and $eB$. 

All these parameter values fall within the range of the standard NJL model, and the percentage change is not much as compared to the ones calculated by solving the relevant expressions. To go to the heavier pions, as mentioned, we use the relevant observable and find the $m$ to attain a particular value of $m_\pi$ while we keep the values of $G_{0}$ and $\Lambda$ fixed to their values, $G_{0}=4.5\; {\rm GeV}^{-2}$ and $\Lambda=650$ MeV. The fitted numbers are mentioned in Table~\ref{tab:parameters}.

\begin{figure}[h!]
  \includegraphics[scale=0.4]{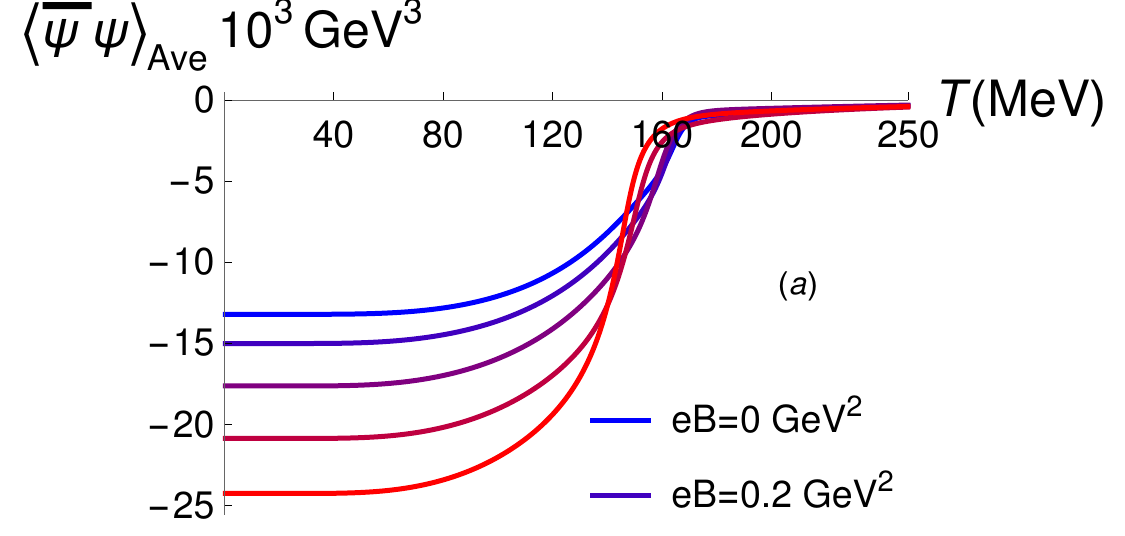}
  \includegraphics[scale=0.4]{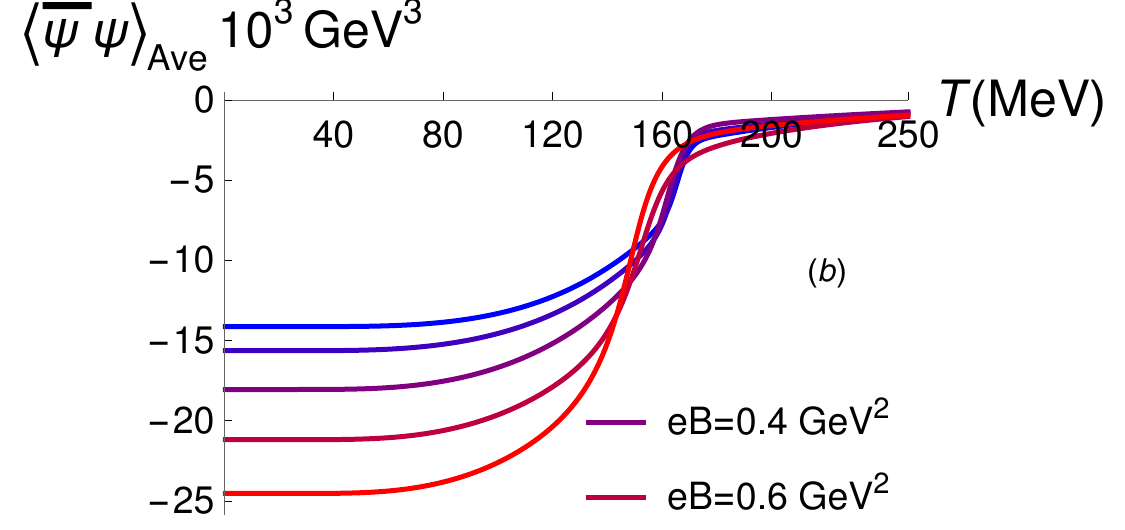}\\
  \includegraphics[scale=0.4]{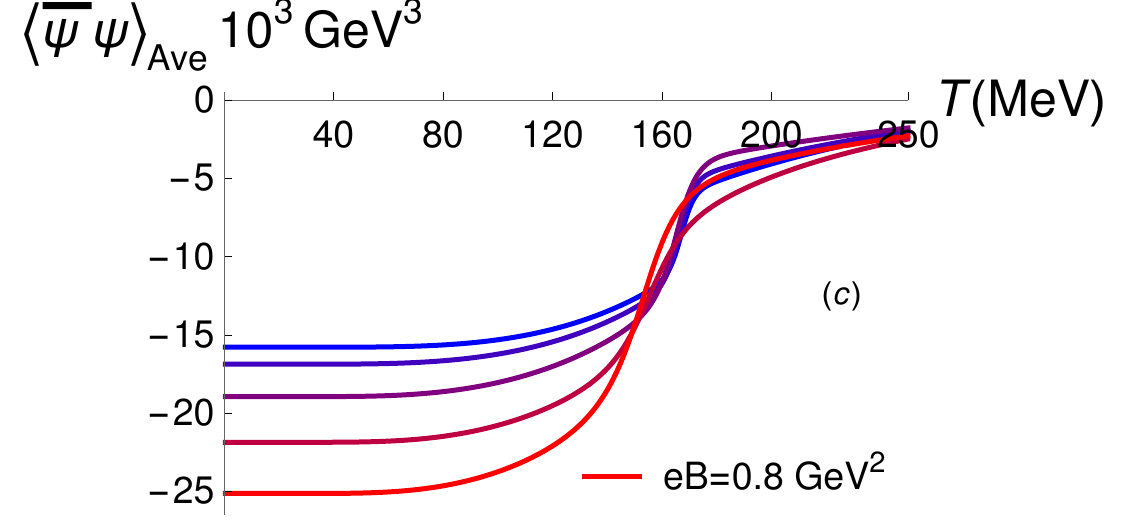}
  \includegraphics[scale=0.4]{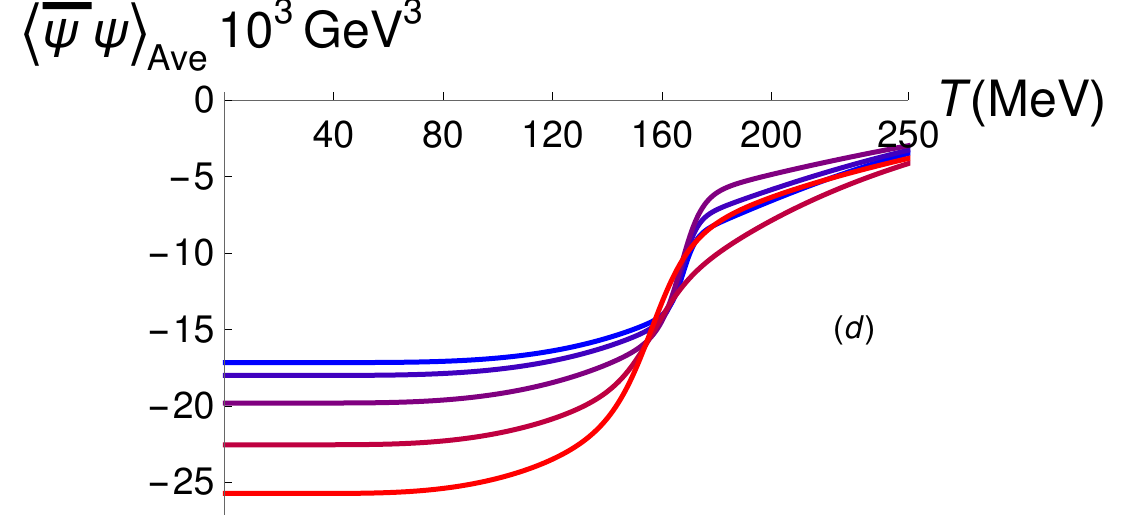}
  \caption{Temperature dependence of the condensate average as a function of $T$ for different $m_\pi$-values in a $2$-flavor local NJL model. For notations and legends, the readers are referred to Fig.~\ref{fig:cond_ave}. Note that here $eB$ reaches up to $0.8\,{\rm GeV}^2$.}
  \label{fig:cond_ave_2f}
\end{figure}
Once the current quark masses are found for a given value of pion mass, we follow the same strategies as for the nonlocal model. We calculate the condensate average for these different pion masses as a function of $T$ for different values of $eB$. The plots are shown in Fig.~\ref{fig:cond_ave_2f}. The first panel expectedly matches with the result of Ref.~\cite{Farias:2016gmy}. As we increase the pion mass, the IMC effect seems to go away for some values of $eB$. At the highest values of $m_\pi$ we investigated, it seems that the IMC effect is gone except for some higher values of $eB$. However, to really know whether there is still some hint of IMC or not, we need to calculate the condensate-average difference, i.e., the change in the condensate average due to the presence of the magnetic field, for which we have dedicated a separate section later. It is difficult to make clear statements on the fate of the IMC effect with increasing values of $m_\pi$ from this Fig.~\ref{fig:cond_ave_2f}. This is in contrast with our observation from the $2$-flavor nonlocal model.

\begin{figure}
    \centering
    \includegraphics[scale=0.37]{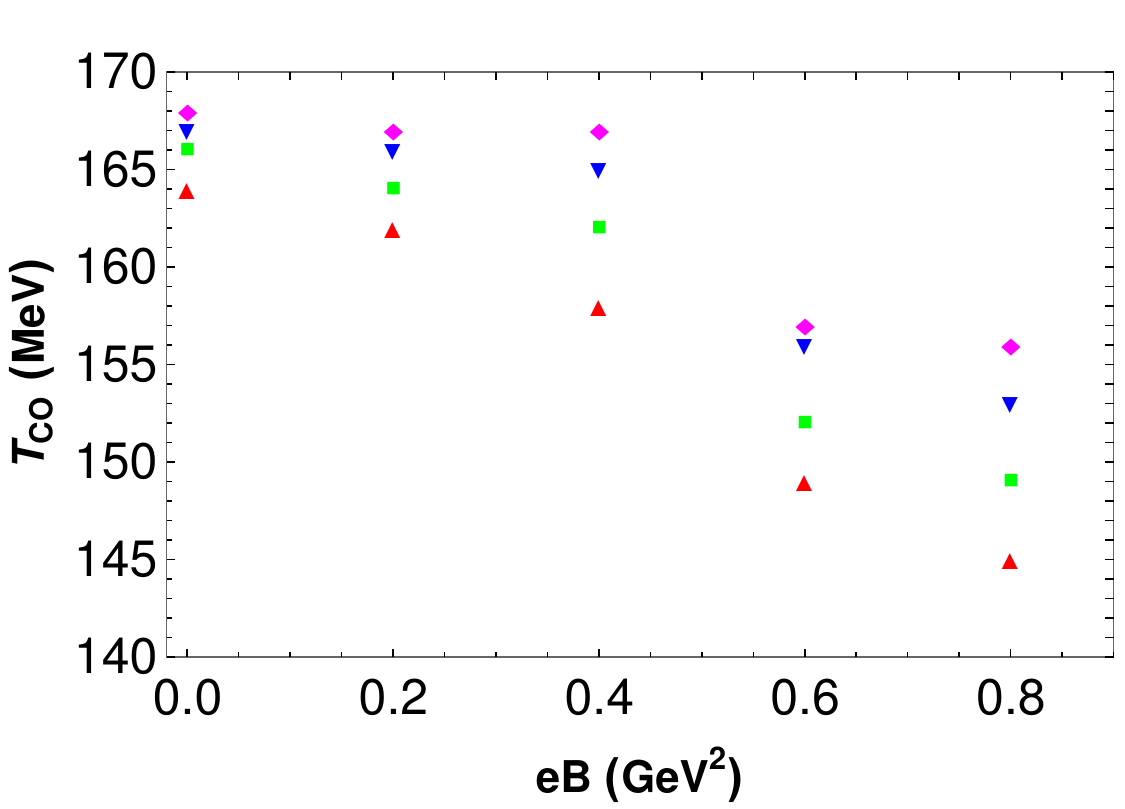}
    \includegraphics[scale=0.55]{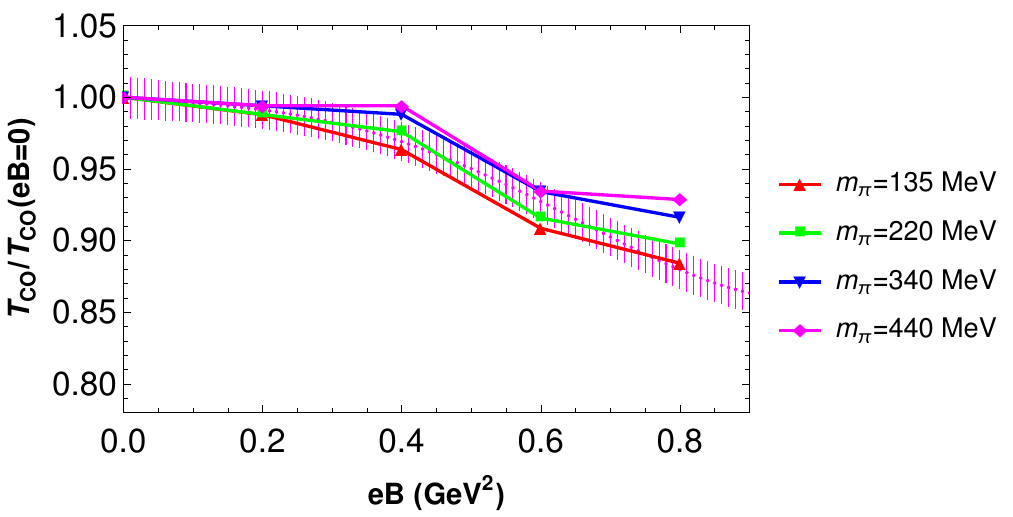}
    \caption{Left: Crossover temperatures as a function of magnetic field for different values of $m_\pi$ for a $2$-flavor local model. Right: The same plot with crossover temperatures scaled with respective zero $eB$-values and compared with LQCD result~\cite{Bali:2012zg} for physical pions (the magenta band). The lines are drawn to guide the eyes.}
    \label{fig:TcVsMpi_local_2f}
\end{figure}
Next, in Fig.~\ref{fig:TcVsMpi_local_2f}, we show the phase diagrams obtained in this model setup. In the left panel, we plot $T_{\text{CO}}$ calculated for different values of $m_\pi$ with increasing $eB$. Though the decreasing trend of $T_{\text{CO}}$ is present for all values of $m_\pi$, the decrease is not regular, and there appears a jump around the $eB$-values $0.4-0.6$ GeV$^{2}$. This can be understood by looking at the behavior of the running of the coupling constant, as discussed at the end of the result section.

In the right panel, we have the same plot scaled by $eB=0$ transition temperatures and compared with the LQCD result. Though scaling makes the result appear relatively smoother, however, the irregular pattern remains for both increasing values of $m_\pi$ and $eB$. Thus, though with the fitting, the model successfully captured the IMC effect and the decreasing behavior $T_{\text{CO}}$, it does not work that efficiently when we try to utilize it beyond the physical point.

\subsubsection{$eB$ dependent coupling constant}
\label{sssec:coup_cons_eB}
Along with the Ref.~\cite{Farias:2014eca}, there came another treatment of the NJL model, which could successfully capture the IMC effect~\cite{Ferreira:2014kpa}. It also uses the same idea of the running of the coupling constant. However, there are some differences. First of all, this is a $2+1$-flavor calculation, and the fitted coupling constant depends only on the external field \footnote{There is another LQCD motivated Polyakov loop extended $2$-flavor NJL study that captures the IMC effect with only an $eB$ dependent coupling constant~\cite{Endrodi:2019whh}.}. The fitting is done to reproduce the chiral $T_{\text{CO}}$ determined by LQCD study~\cite{Bali:2011qj} at different values of $eB$. This is another well-known method to capture the IMC effect in a local version and has been so far extensively used by the community. Thus, we use it here to test its applicability beyond the physical point.

As mentioned earlier, the parameter fitting is more involved here since there is now strange quark in the system. To go beyond the physical point, we keep the ratio of strange to light quark ratio to its physical value, $25.58$, following the arguments in Refs~\cite{Fukushima:2008wg,DElia:2018xwo}. Thus, to attain a higher value of $m_\pi$, one needs to not only change the light quark mass but also adjust the strange quark mass to satisfy the aforementioned ratio. We follow this procedure while solving the relevant observables equation consistently keeping both the coupling constants, $G_{1}$ and $G_{2}$, and the cut-off, $\Lambda$ fixed: $G_{1}=3.67/\Lambda^2$, $G_{2}=12.36/\Lambda^5$ and $\Lambda=602.3$ MeV. Accordingly, we find the light quark masses and the corresponding $m_\pi$ values as given in Table~\ref{tab:parameters}.

\begin{figure}[h!]
  \includegraphics[scale=0.4]{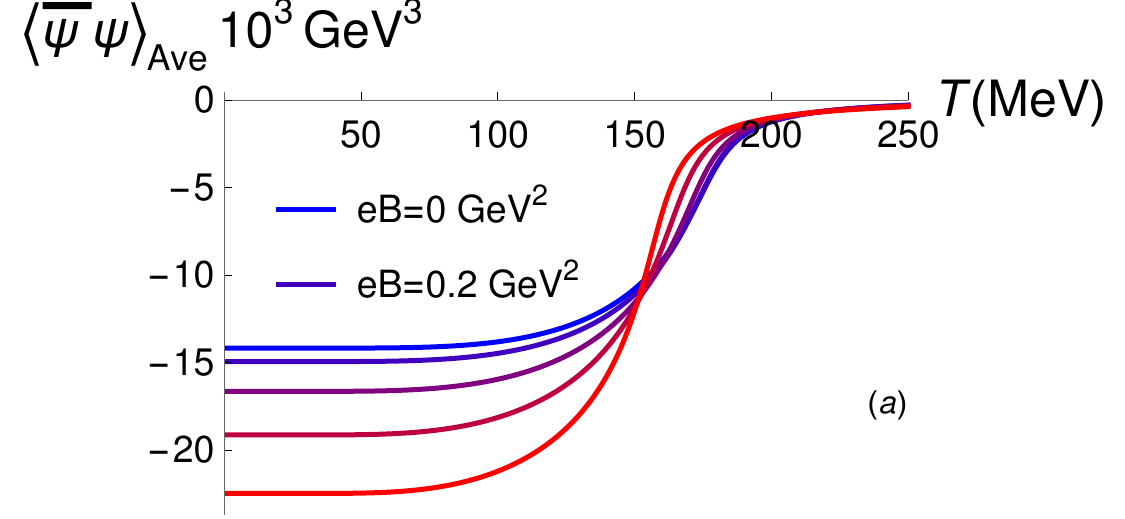}
  \includegraphics[scale=0.4]{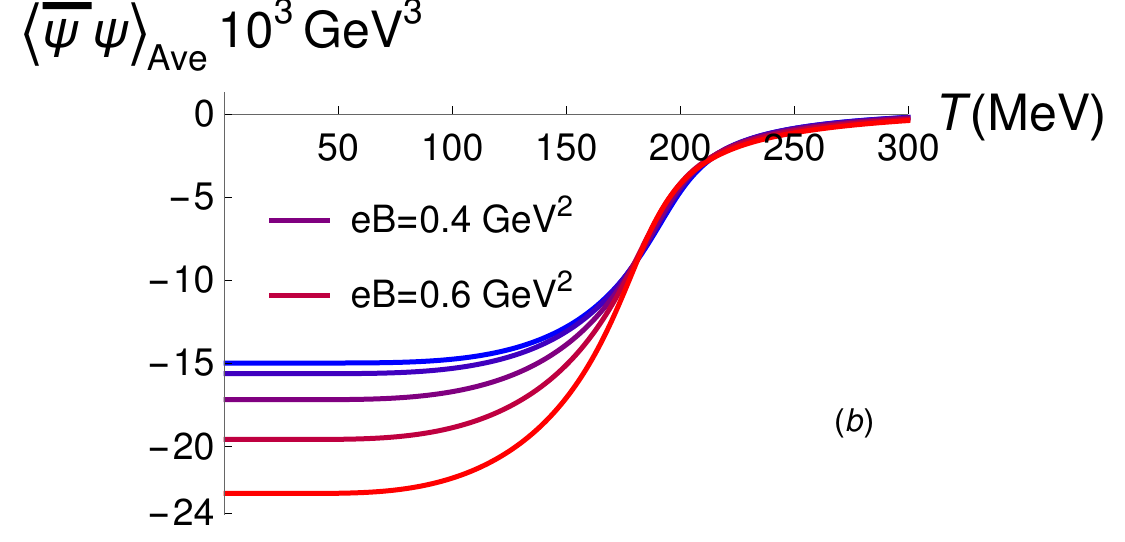}
  \includegraphics[scale=0.4]{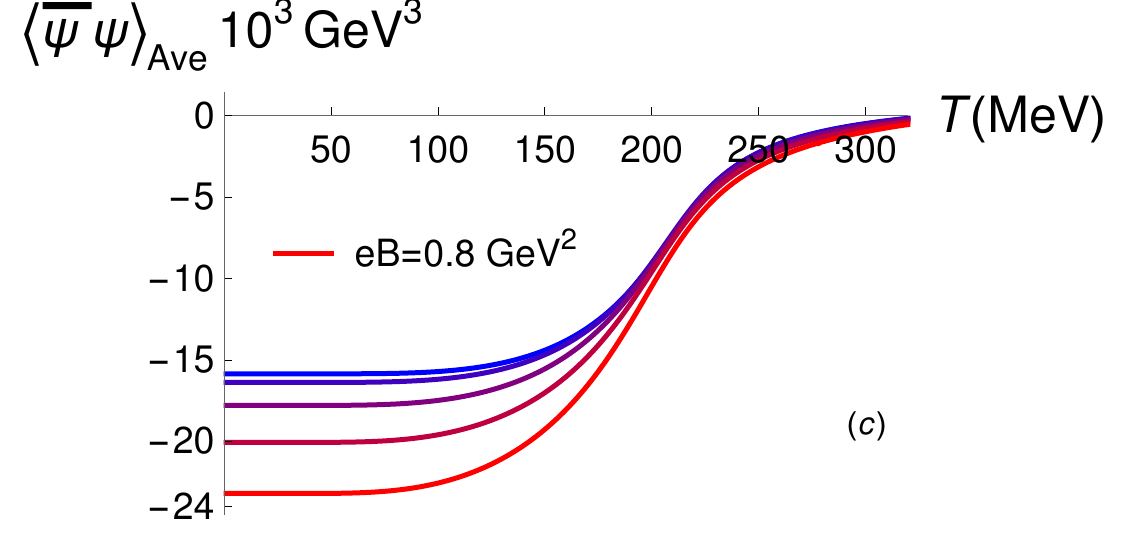}
  \includegraphics[scale=0.4]{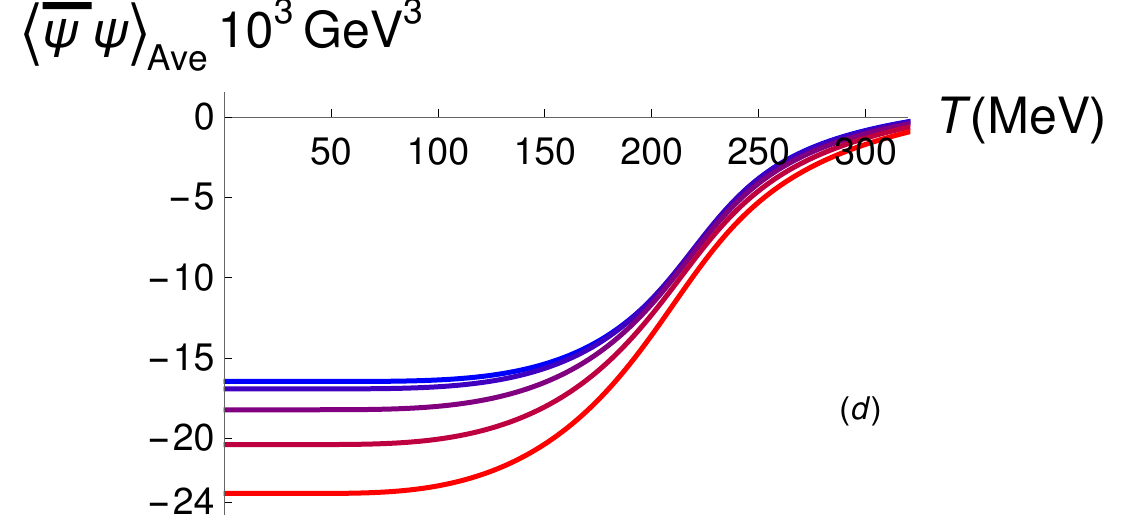}
  \caption{Temperature dependence of the condensate average as a function of $T$ for different $m_\pi$ in a $2+1$-flavor local NJL model. For notations and legends, the readers are referred to Fig.~\ref{fig:cond_ave}. Here $eB$ has the maximum strength of $0.8\,{\rm GeV}^2$.}
  \label{fig:cond_ave_2plus1}
\end{figure}
With the mentioned light quark masses for different values of $m_\pi$ we investigated the condensate average in Fig.~\ref{fig:cond_ave_2plus1}. In panel (a), we have the plot for the physical point with a clear demonstration of the IMC effect around the phase transition region. With $m_\pi=220$ MeV, the IMC effect is reduced but not completely wiped out. As we increase $m_\pi$ further, the IMC effect appears to be totally eliminated in panels (c) and (d). However, to really demand the elimination of the IMC effect for all values of $eB$, one needs to look into the change in the condensate average due to the presence of the magnetic field, which we discuss in the next subsection. However, we can remark that the fate of the IMC effect with increasing $m_\pi$ behaves qualitatively similar to that found in the LQCD study~\cite{DElia:2018xwo}.

\begin{figure}
    \centering
    \includegraphics[scale=0.37]{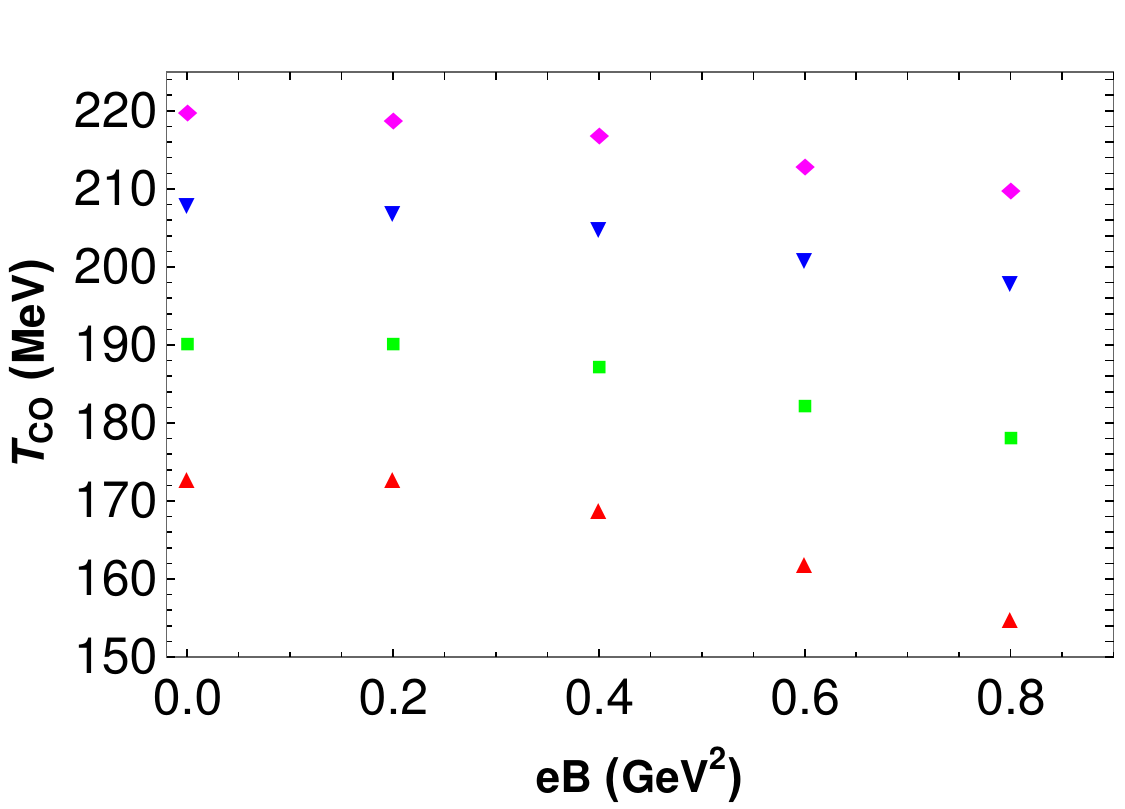}
    \includegraphics[scale=0.55]{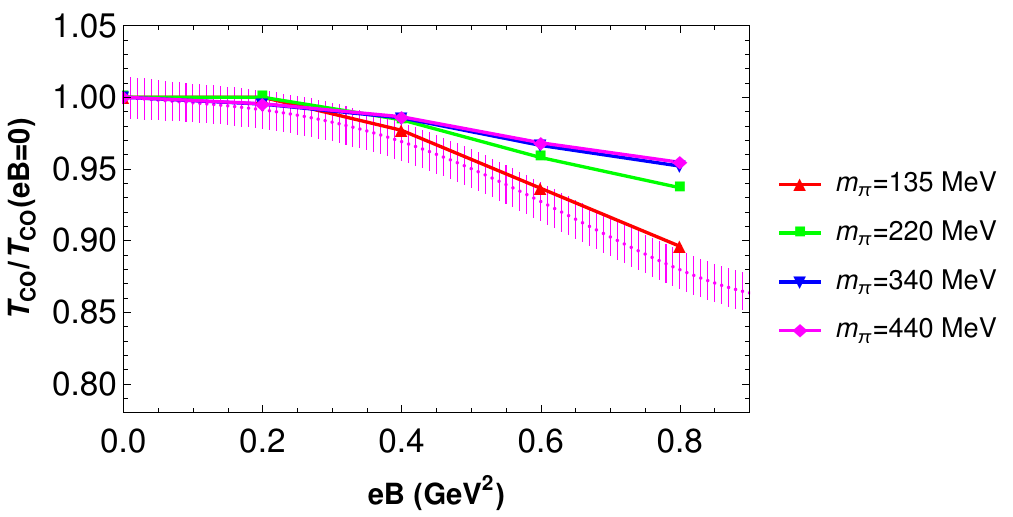}
    \caption{Left: Crossover temperatures as a function of $eB$ for different values of $m_\pi$ in a $2+1$-flavor local model. Right: The same plot with crossover temperatures scaled with respective zero $eB$-values and compared with LQCD result~\cite{Bali:2012zg} for physical pions (the magenta band). The lines between the points exist solely to guide the eyes.}
    \label{fig:TcVsMpi_local_2plus1f}
\end{figure}
As we look at the plot for crossover temperature as a function of $eB$ for different values of $m_\pi$ in the left panel of Fig.~\ref{fig:TcVsMpi_local_2plus1f}, we observe that the model also successfully captures the behavior of decreasing $T_{\text{CO}}$ found in LQCD. Here, the behavior is smooth as opposed to what we found for the $2$-flavor local model in Fig.~\ref{fig:TcVsMpi_local_2f}. In the right panel, we display the phase diagram with scaled transition temperatures and compare it with the LQCD data. The red points for the physical pion mass, as anticipated, match with the data well, as the model was fit to match with these data. With higher values of $m_\pi$, the decreasing trend of the scaled transition temperature goes up for higher values of $eB$. This is qualitatively similar to the LQCD result. Thus, as compared to the $2$-flavor local model, the $2+1$-flavor reproduces more reliably results beyond the physical pion mass.

\subsection{Condensate-average differences with respect to its value at $B=0$}
\label{ssec:cond_diff}
\begin{figure}[h!]
  \includegraphics[scale=0.8]{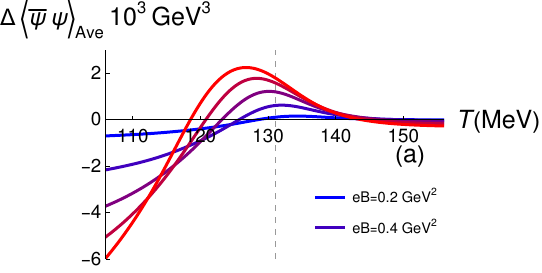}
  \includegraphics[scale=0.8]{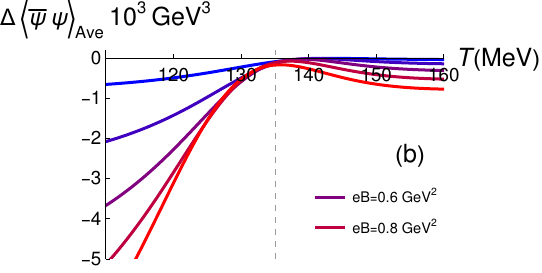}\\
  \includegraphics[scale=0.8]{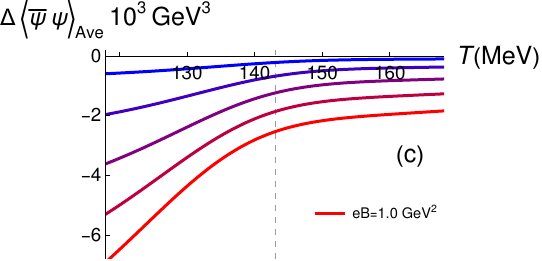}
  \includegraphics[scale=0.8]{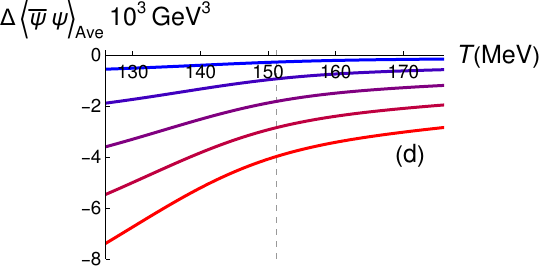}
  \caption{(Nonlocal $2$-flavor): The condensate-average difference plotted as a function of temperature for different values of $eB$, with increasing strength from blue to red as denoted by the legends. Panels (a), (b), (c), and (d) represent $m_\pi$-values of $135$, $220$, $340$, and $440$, respectively. The condensate values are scaled with a factor of $10^3$. The gray dashed line represents the $T_{\rm CO}$ in the respective model at $eB=0$.}
  \label{fig:dcond_avedB}
\end{figure}
In an effort to understand the correlation between the IMC effect around the crossover temperature and the decreasing $T_{\text{CO}}$, we calculate the condensate-average difference, $\Delta\langle\bar\psi\psi\rangle_{\text{Ave}}$ and investigate it for all the above-mentioned frameworks. It is defined as the subtraction of the condensate average for $eB=0$ from the ones with non-zero $eB$,
\begin{align}
\Delta\langle\bar\psi\psi\rangle_{\text{Ave}} (eB,T,m)=\langle\bar\psi\psi\rangle_{\text{Ave}} (eB,T,m)-\langle\bar\psi\psi\rangle_{\text{Ave}} (0,T,m).
\label{eq:cond_ave_diff}
\end{align}
If the IMC effect persists for a given value of $eB$ with a certain $m_\pi$, then $\Delta\langle\bar\psi\psi\rangle_{\text{Ave}}$ becomes positive. 

In Fig.~\ref{fig:dcond_avedB}, the condensate-average differences are shown in the nonlocal framework. It is evident that the IMC effects are present in panel (a) for the physical $m_\pi$, which corroborates with panel (a) in Fig.~\ref{fig:cond_ave}. From panel (b) of Fig.~\ref{fig:cond_ave}, it was not possible to make a definite comment on the elimination of the IMC effect for all values of $eB$. However, panel (b) in Fig.~\ref{fig:dcond_avedB} demonstrates that the IMC effect is indeed eliminated at $m_\pi=220$ in the nonlocal model. The plots from the other panels in Fig.~\ref{fig:dcond_avedB} are consistent with our understanding from the respective pion mass plots in Fig.~\ref{fig:cond_ave}. The gray dashed lines are the zero $eB$ crossover temperature at relevant $m_\pi$-values. They are drawn to illustrate the temperature range where the IMC effect appears.

\begin{figure}[h!]
  \includegraphics[scale=0.4]{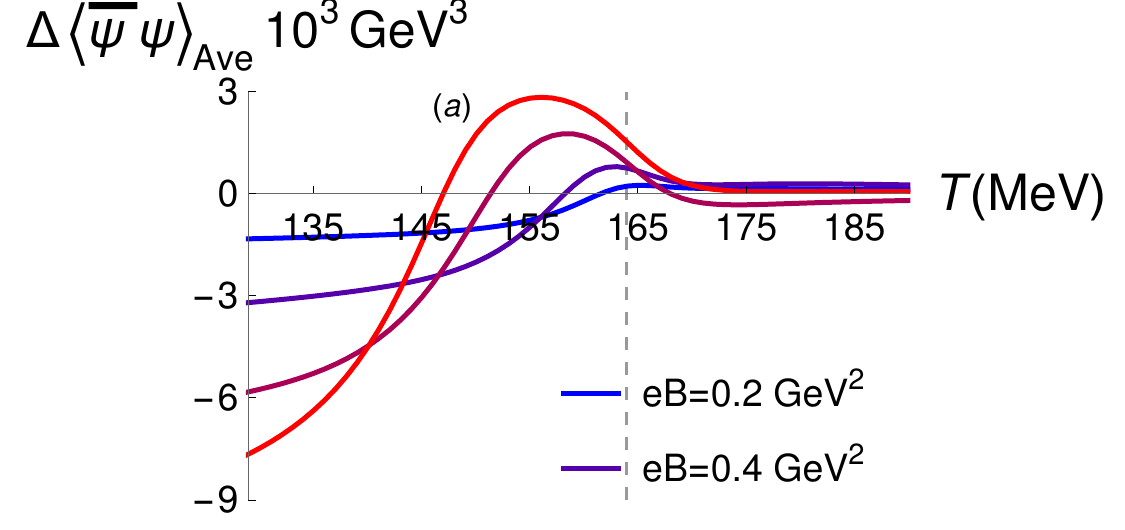}
  \includegraphics[scale=0.4]{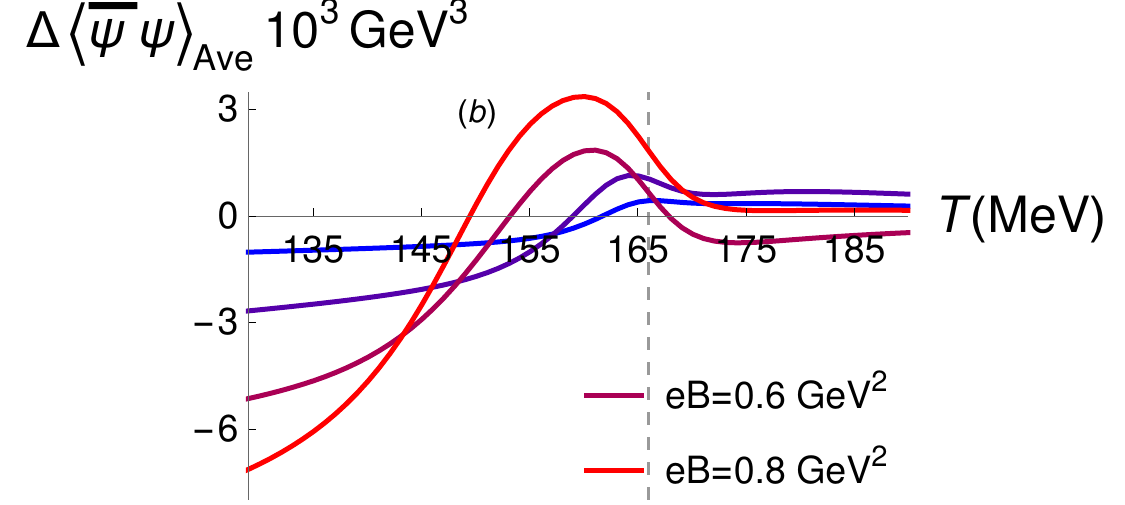}\\
  \includegraphics[scale=0.4]{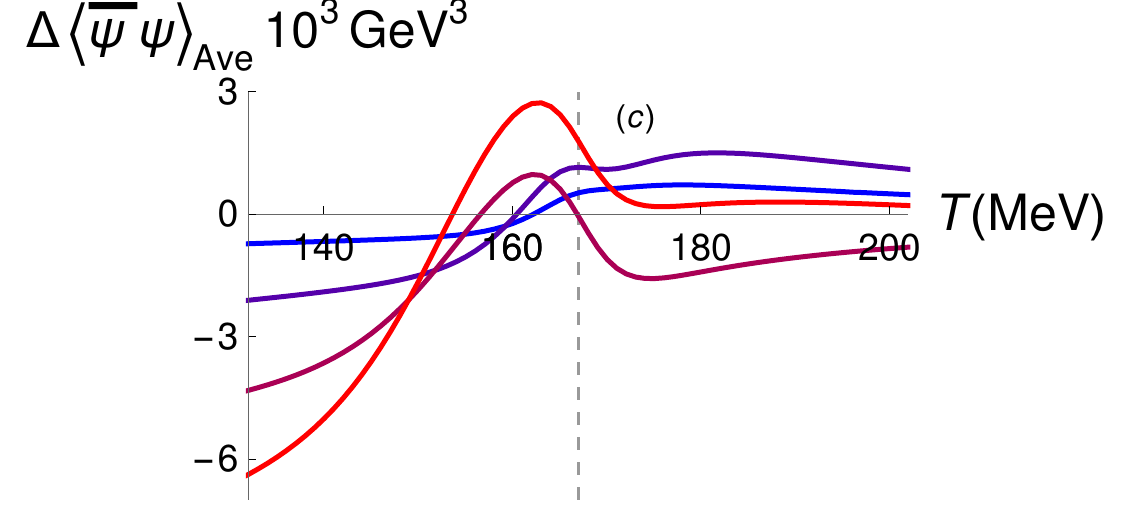}
  \includegraphics[scale=0.4]{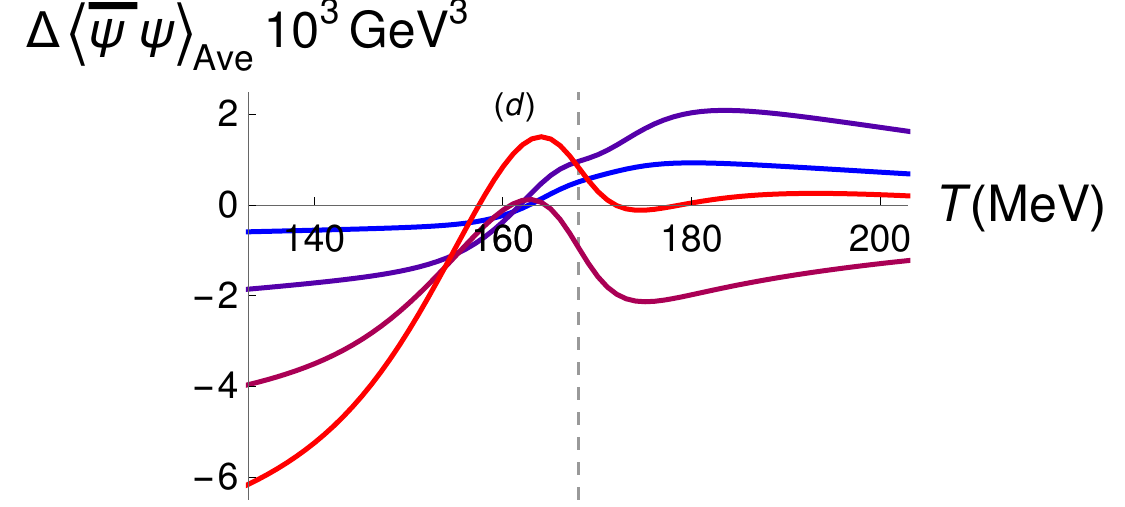}
  \caption{(Local $2$-flavor): The condensate-average difference plotted as a function of temperature for different values of $eB$, with increasing strength from blue to red as denoted by the legends. The notations and legends are the same as in Fig.~\ref{fig:dcond_avedB}. The gray dashed line represents the $T_{\rm CO}$ in the respective model at $eB=0$.}
  \label{fig:dcond_avedB_local2f}
\end{figure}
Now, we move to the local $2$-flavor NJL model. The plots for $\Delta\langle\bar\psi\psi\rangle_{\rm Ave}$ for different $m_\pi$'s are shown in Fig.~\ref{fig:dcond_avedB_local2f}. The results for the condensate average and the phase diagram beyond physical $m_\pi$ did not look very encouraging in this framework. This is again reflected in the plots of $\Delta\langle\bar\psi\psi\rangle_{\rm Ave}$. The IMC effect is present for all values of $m_\pi$ considered here. Also, the relative trend of $\Delta\langle\bar{\psi}\psi\rangle_{\rm Ave}$ for heavier than physical pions does not show a consistent pattern as we change $eB$. The description even worsens and becomes random after the restoration of the symmetry.

\begin{figure}[h!]
  \includegraphics[scale=0.45]{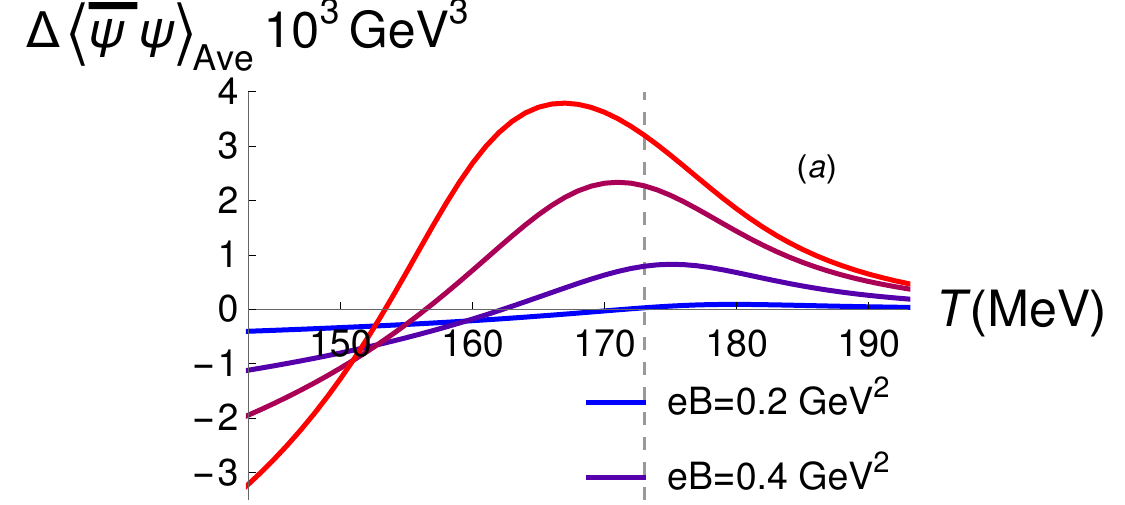}
  \includegraphics[scale=0.45]{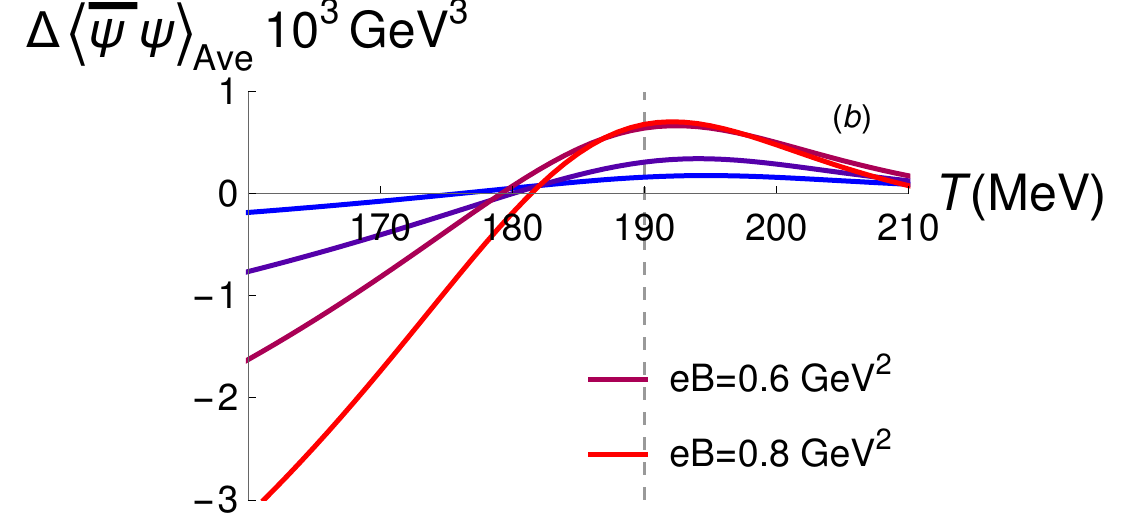}\\
  \includegraphics[scale=0.45]{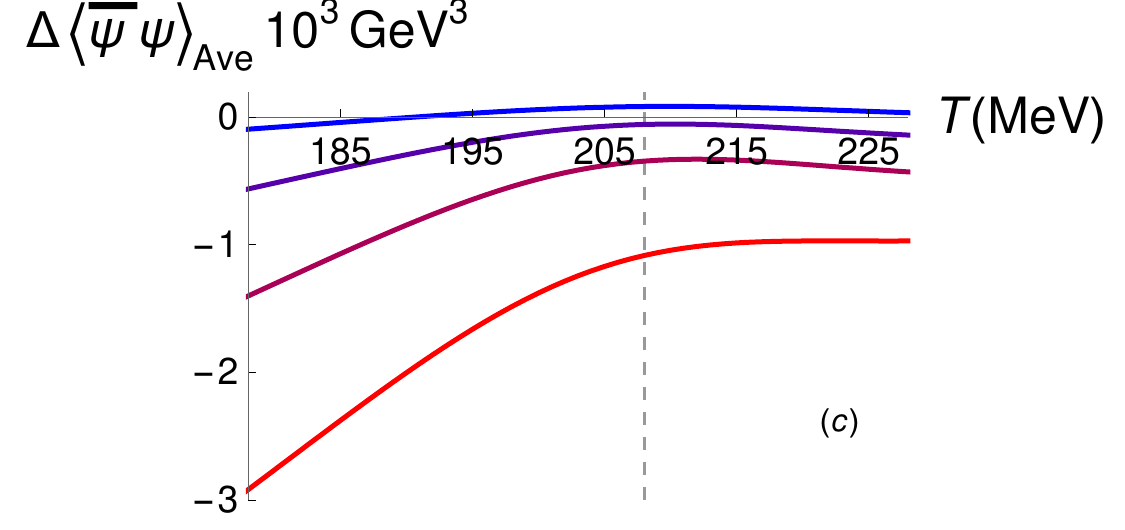}
  \includegraphics[scale=0.45]{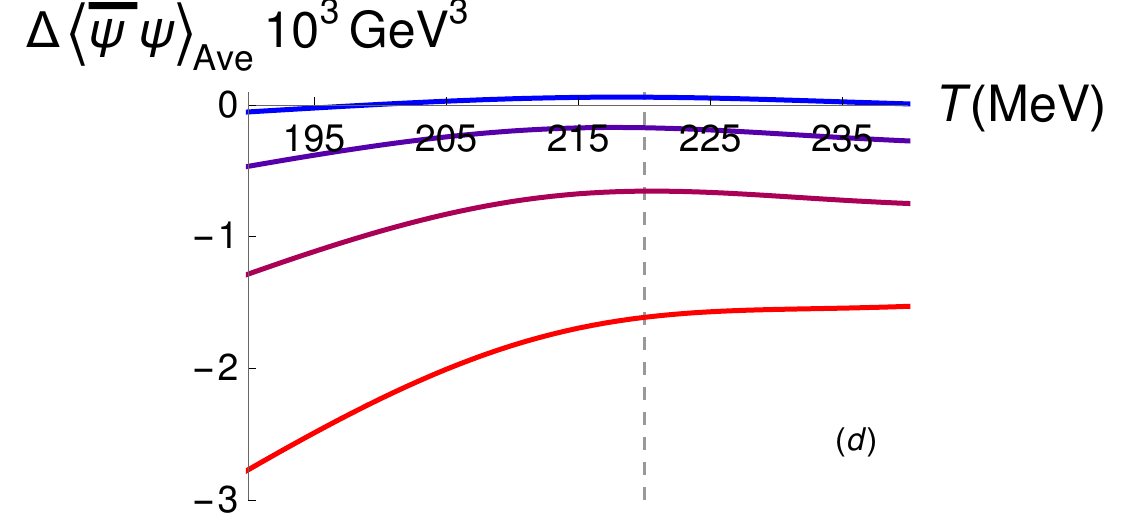}
  \caption{(Local $2+1$-flavor): The condensate-average difference plotted as a function of temperature for different values of $eB$ with increasing strength from blue to red as denoted by the legends. The notations and legends are the same as in Fig.~\ref{fig:dcond_avedB}. The gray dashed line represents the $T_{\rm CO}$ in the respective model at $eB=0$.}
  \label{fig:dcond_avedB_local2plus1f}
\end{figure}
The plots for the condensate-average differences for the $2+1$-flavor local NJL model are shown in Fig.~\ref{fig:dcond_avedB_local2plus1f}. Panels (a) and (b) confirm the presence of the IMC effect, thus corroborating the respective panels of Fig.~\ref{fig:cond_ave_2plus1}. In this framework, the IMC effect is present for $m_\pi=220$ MeV, although with a non-smooth behavior for different $eB$'s. To compare, we have shown the results in the nonlocal framework at $m_\pi=180$ MeV in the appendix~\ref{sec:app_non_local}, which shows a gradual enhancement with increasing $eB$. With even higher $m_\pi$ values in Fig.~\ref{fig:dcond_avedB_local2plus1f}, the IMC effect disappears for all the values of $eB$, except $eB=0.2\,\text{GeV}^2$. For this value of $eB$, the IMC effect persists for all $m_\pi$-values we tested. This observation is not obvious from panels (c) and (d) of Fig.~\ref{fig:cond_ave_2plus1}. In summary, the description of the condensate for heavier pion mass QCD medium in the $2+1$ framework is qualitatively satisfactory; however, it shows IMC for $eB=0.2\,{\text{GeV}}^2$ for $m_\pi=440$MeV, contrary to LQCD results~\cite{DElia:2018xwo}. 

\subsection{The pion mass beyond which the IMC effect disappears.}
\label{ssec:crit_pion}
\begin{figure}[h!]
  \includegraphics[scale=0.9]{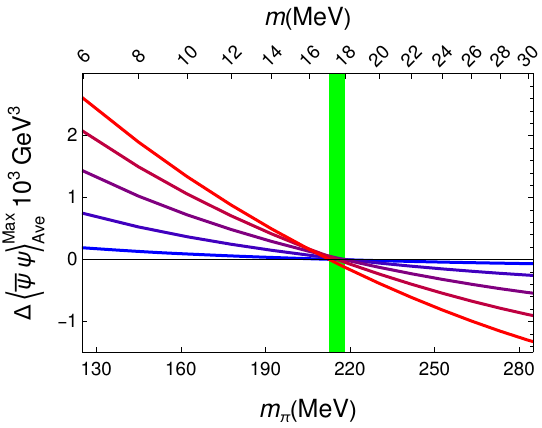}
  \caption{The pion mass above which the IMC effect disappears for various $eB$-values. The strength of the magnetic field increases from $0.2$ $\text{GeV}^2$ (blue) to $1.0$ $\text{GeV}^2$ (red) in steps of $0.2$ $\text{GeV}^2$.}
  \label{fig:crit_pion}
\end{figure}
In this section, inspired by the LQCD study~\cite{Endrodi:2019zrl}, we obtain the pion mass above which the IMC effect around the crossover temperature ceases to exist, which we call the maximum pion mass $(m_{\pi}^{\text{max}})$. Ref.~\cite{Endrodi:2019zrl} is a $2+1$-flavor study which is done by fixing the strange quark mass to its physical value. This is in contrast to Ref.~\cite{DElia:2018xwo} in which the strange to light quark mass ratio is kept fixed at a physical value. We followed the latter approach while performing the $2+1$-flavor local framework in Sec.~\ref{ssec:loc_mod}.

Out of the three frameworks that we tested, the nonlocal $2$-flavor model is one where we can easily estimate such a quantity. As demonstrated in the previous section, we cannot draw clear conclusions about the fate of the IMC effect with increasing $m_\pi$ in the $2$-flavor local model. This is also true for the $2+1$-flavor local model, in which the IMC effect persists at $eB=0.2\,{\rm GeV}^2$ for all values of the pion mass that are tested.

In Fig.~\ref{fig:crit_pion}, we have plotted the maximum value of the condensate-average difference [Eq.~\eqref{eq:cond_ave_diff}] as a function of the (vacuum) pion mass for different magnetic fields in the non-local model. For convenience, we have quoted the corresponding current quark masses on the upper axis. For a fixed magnetic field, $m_{\pi}^{\text{max}}$ is the vacuum pion mass for which this condensate-average difference becomes zero. Ref.~\cite{Endrodi:2019zrl}, which is a $2+1$-flavor LQCD study, has obtained the maximum mass for a single magnetic field $eB=0.6\ \text{GeV}^{2}$ which is $497(4)$ MeV. On the other hand, our model's simplicity allowed us to obtain the maximum mass for the whole range of $eB$ we explored. We found that the maximum value has a little $eB$ dependence with a spread of $1.4\%$ around the mean value of $m_{\pi}^{\text{max}}=215\ \text{MeV}$. This value is much smaller as compared to the corresponding value in LQCD. However, we emphasize that qualitatively the models can capture the LQCD result beyond the physical point, and it is encouraging to find a maximum pion mass similar to the LQCD study within the model framework as well.

\subsection{Characteristics of the coupling constants}
\label{ssec:coup_cons}
In this sub-section, we describe the behavior of the coupling constant in different frameworks in the presence of the magnetic field. In the local NJL model, one assumes the coupling strength to be constant while obtaining the free energy. On the other hand, in the nonlocal model, the coupling constant runs with the momentum~\cite{Hell:2008cc}. These features arising from the construction of the respective models play a crucial role in obtaining the IMC effect. In a nonlocal model, the IMC effect is captured automatically. On the other hand, in the local models, one needs to introduce a parameterized running of the coupling constant to reproduce the IMC effect.

\begin{figure}[h!]
  \includegraphics[scale=0.54]{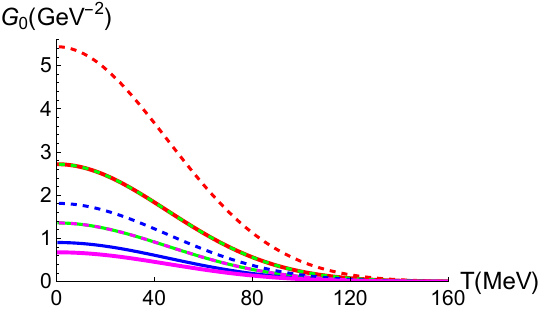}
  \includegraphics[scale=0.42]{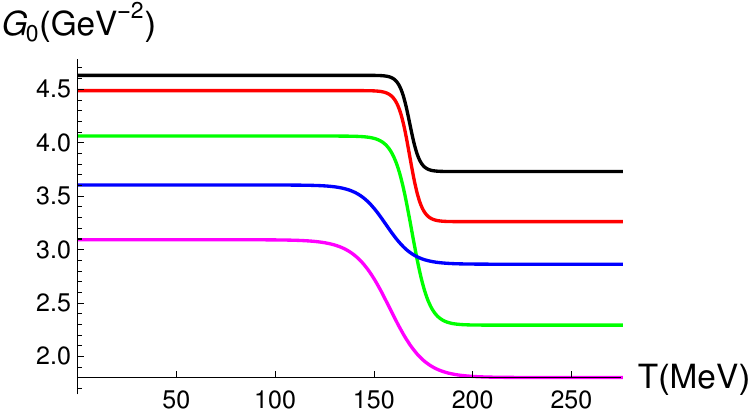}
  \includegraphics[scale=0.42]{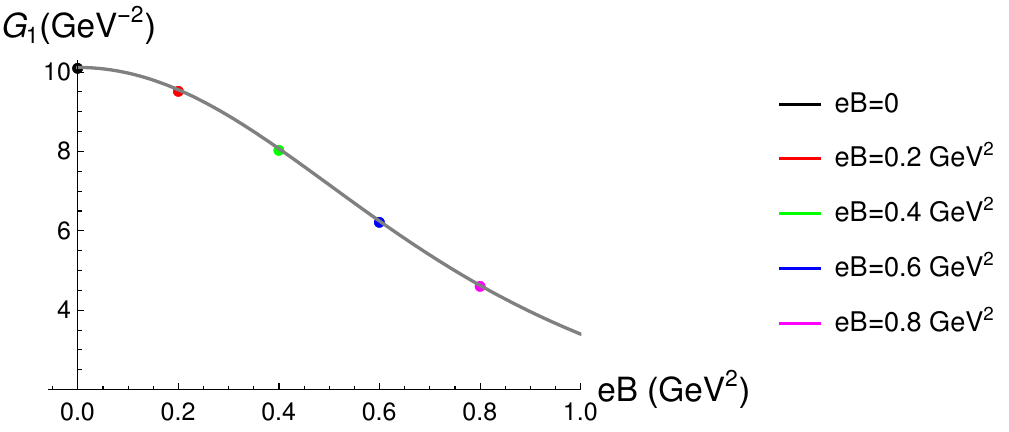}
  \caption{The effective coupling as a function of $T$ and $eB$. The left panel is for the nonlocal model, with the continuous and dashed lines for $u$ and $d$ flavor, respectively. The middle panel represents effective coupling in the $2$-flavor local case~\cite{Farias:2016gmy}. The right panel represents the same in the $2+1$-flavor local model.}
  \label{fig:GvseB}
\end{figure}

So far, our results suggest that the same running enables models to describe the effects of heavier current quark masses without introducing additional parameters. However, while investigating the details, we find out differences among these frameworks, which can be understood by looking at the behavior of the coupling constant in different frameworks as a function of $T$ and $eB$ (Fig.~\ref{fig:GvseB}).

The middle panel of Fig.~\ref{fig:GvseB} represents the effective coupling as a function of $T$ and $eB$ in the $2$-flavor local model. It is obtained using Eq.~\eqref{eq:coupling_2} with the parameter values taken from Ref.~\cite{Bandyopadhyay:2023lvk}. The effective coupling in a local $2+1$-flavor model as a function of $eB$ is shown in the right panel of the same figure. It is obtained from Eq.~\eqref{eq:coupling_2plus1} with appropriate parameters taken from Ref.~\cite{Ferreira:2014kpa}.

The effective coupling in the $2$-flavor nonlocal model is shown in the left panel of Fig.~\ref{fig:GvseB}. Unlike its local counterparts, it is not straightforward to interpret. We identify the effective coupling as the coefficient of $\bar{\sigma}$ in the effective mass relation as given in Eq.~\eqref{eq:eff_mass_B}. Apart from an explicit flavor dependence, this effective coupling depends on the Landau level, spin polarization, momentum, temperature, and Matsubara frequencies of the quark. For simplicity, we have summed up the landau levels and the polarization. On the other hand, the momentum and temperature have a simple exponential form; hence, we have presented the interaction strength for the lowest Matsubara frequency with zero momentum mode.

Thus, apart from the major difference in the implementation of varying coupling, there is another difference that can be important while dealing with magnetic fields: we find that the effective coupling is independent of the flavor in the local models. In other words, the constituent masses for different flavors have an explicit dependency on $eB$ and/or $T$ through the coupling $G(eB)$. However, this dependency is blind to the charges of the associated quark flavor. This is evident from the middle and the right panels of Fig.~\ref{fig:GvseB} as well as from Eqs.~\eqref{eq:coupling_2} and~\eqref{eq:coupling_2plus1}. Contrary to that, a nonlocal version takes into account the strength with which a particular quark flavor interacts with the magnetic field (left panel in Fig.~\ref{fig:GvseB}).

By looking at the middle panel, we can explain the irregular behavior of the IMC effect found in Fig.~\ref{fig:dcond_avedB_local2f}. The running of the coupling has a sudden drop in strength around the $T_{\rm CO}$ for all $eB$-values. There is also a crossing of the strength of the coupling between $0.4$ and $0.6$ ${\rm GeV}^2$. This explains the irregular trend in $T_{\rm CO}$'s for all values of $m_\pi$. On the other hand, panel (b) in Fig.~\ref{fig:dcond_avedB_local2plus1f} produces IMC effect around the crossover temperature at $m_\pi=220$ MeV with an irregular strength for different $eB$'s. This happens due to the flavor-independent nature of the running of the coupling, as shown in the right panel of Fig.~\ref{fig:GvseB}. In fact, this flavor independence of the coupling is the reason that the IMC effect cannot be eliminated for the lowest strength of the $eB$ [panels (c) and (d) in Fig.~\ref{fig:dcond_avedB_local2plus1f}] for higher pion masses.

On the contrary, the results in the nonlocal model look consistent throughout. This can be attributed to the flavor-dependent smooth running of the coupling, as shown in the left panel in Fig.~\ref{fig:GvseB}. In fact, this is the reason that in this model, the observed IMC effect in a heavier pion mass ($m_\pi=180$ MeV) behaves consistently as described in the appendix~\ref{sec:app_non_local}. This is contrary to panel (b) in Fig.~\ref{fig:dcond_avedB_local2plus1f}, where the behavior is irregular for different $eB$-values. 

To conclude this section, it is evident from the left panel of Fig.~\ref{fig:GvseB} that the effective coupling decreases with increasing magnetic field, with a stronger effect for the $u$ quark than for the $d$ quark at a given $eB$. The interaction strength for the $u$ quark in a magnetic field of strength $eB$ is equal to that for the $d$ quark in a field of strength $2eB$. The behavior of the individual light quark condensate in the presence of $eB$ with increasing $m_\pi$-values is detailed in the appendix~\ref{sec:app_imc_indv}.

%%%%%%%%%%%%%%%%%%%%%%%%%%%%%%%%%%%%%%%%%%%%%%%%%%%%%%%%%%%%%%%%%%%%%%%%
\section{Summary and conclusion}
\label{sec:summary}
%%%%%%%%%%%%%%%%%%%%%%%%%%%%%%%%%%%%%%%%%%%%%%%%%%%%%%%%%%%%%%%%%%%%%%%%
We have explored the effect of the varying pion mass on the chiral crossover line of a magnetized QCD medium within an effective model framework. It is now known that to obtain the inverse magnetic catalysis (IMC) around the crossover temperature ($T_{\text{CO}}$), as first observed in lattice QCD (LQCD) simulation, the running of the effective coupling plays the most crucial role. Such running of the coupling has been implemented in different ways in effective models. Following a recent LQCD result~\cite{DElia:2018xwo}, we have explored these models with larger pion mass beyond the physical point in the presence of magnetic fields, which, to the best of our knowledge, is the first in an effective model scenario.

We utilize the widely used Nambu\textendash Jona-Lasinio (NJL) model and study both its local and nonlocal versions. We use the nonlocal model constructed with two light quark flavors. In this model, the coupling runs with the momentum. This feature, when incorporated appropriately, naturally bears out the IMC effect in the presence of an external magnetic field. For a local interaction, the coupling constant is considered to be independent of the momentum up to the cutoff scale of the theory (in this case, the 3-momentum cutoff). But to achieve the IMC effect, one needs to make it dependent on available thermodynamic parameters, for example, temperature, external magnetic field, etc. In this work, we explored two local model scenarios: one with a temperature and magnetic field-dependent coupling constant, constructed with two light quark flavors. The other one, with only a magnetic field-dependent coupling strength, is constructed with two light and a heavier strange quark. On the other hand, the nonlocal model, constructed with two light flavors of quark, takes into the effects of every thermodynamic parameter present in the theory naturally.

We observe that the IMC effect ceases to appear as we increase the pion mass, but the decreasing trend of the $T_{\rm CO}$ persists. Thus, the NJL model is able to reproduce the observations found in LQCD qualitatively. However, the value of the pion mass at which the IMC effect goes away is lower than what is found in the LQCD study~\cite{DElia:2018xwo}. We also find the pion mass above which the IMC effect around the crossover temperature ceases to exist. In the nonlocal model, the IMC effect disappears around $m_\pi=215$ MeV with a spread of $1.4\%$ due to its magnetic field dependency. The LQCD estimate is $497(4)$ MeV at $eB=0.6\ \text{GeV}^{2}$. Obtaining quantitative agreement with LQCD may require introducing additional mass and $eB$ dependent parameters in the model. The other frameworks are not suitable to perform such an analysis, as the IMC effect is not consistently produced for beyond physical point. For example, in the $2+1$-flavor local NJL model, the IMC effect exists for all values of $m_\pi$ that we tested for $eB=0.2\ \text{GeV}^{2}$. For magnetic filed larger than $0.2\ \text{GeV}^{2}$ it disappears between $m_\pi=220$ and $340$ MeV.   

Although the observations hold for both the nonlocal and local frameworks, for the local version, it is dependent on the way the medium-dependent coupling is implemented in the model. We observe some irregularities at larger pion masses that might stem from the fitting of the parameter at the physical point in the local model. In $2$-flavor local framework, the decreasing trend of $T_{\rm CO}$'s is not smooth as we increase the pion mass. Also, it is not possible to make any conclusive comment there on the status of the IMC effect beyond the physical point. In the $2+1$-flavor local model, the IMC effect for higher $m_\pi$-value ($220$ MeV) shows an irregular pattern for $eB$-dependence and the IMC effect is present for $eB=0.2\,{\rm GeV}^2$ for all the $m_\pi$-values tested. In contrast, the IMC effect for beyond physical $m_\pi$-value shows smooth behavior in the nonlocal model.

We further explored the effective coupling to understand the results. In the local scenario, though the coupling constants are magnetic field dependent, it is blind to the strength (here, the charges of the quarks) at which a particular flavor interacts with the magnetic field. We find that the fitting of the parameters to reproduce LQCD results at the physical point is not enough to capture the physics for heavier pion masses. In that respect, we find that a smooth effective coupling via a form factor in the nonlocal model captures the physics more naturally. This is further analyzed in appendix~\ref{sec:app_imc_indv} by showing the behavior of the individual condensates.

Through the detailed analysis, we tested the reliability of the NJL-like models by going to unphysical regions guided by LQCD observations. We conclude that such simple effective models, when treated consistently, can capture the essence of a complex theory like QCD.\\

%%%%%%%%%%%%%%%%%%%%%%%%%%%%%%%%%%%%%%%%%%%%%%%%%%%%%%%%%%%%%%%%%%%%%%%%

\textbf{Acknowledgment:} The authors would like to thank Gergely Endr\H{o}di and Matteo Giordano for valuable communication. M.S.A. would like to thank Deeptak Biswas for useful discussions. He also acknowledges the Tata Institute of Fundamental Research for providing the facilities and conducive research environment during the initial stages of this work. C.A.I. would like to thank Aritra Bandyopadhyay and Ricardo Luciano Sonego Farias for useful communication. He also acknowledges the support by the State of Hesse within the Research Cluster ELEMENTS with project ID 500/10.006.\\

\begin{appendices}
\label{sec:app}
\section{At $m_\pi=180$ MeV in $2$-flavor nonlocal model}
\label{sec:app_non_local}
\begin{figure}[h!]
  \includegraphics[scale=0.8]{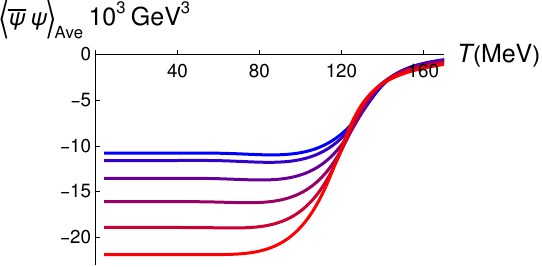}
  \includegraphics[scale=0.8]{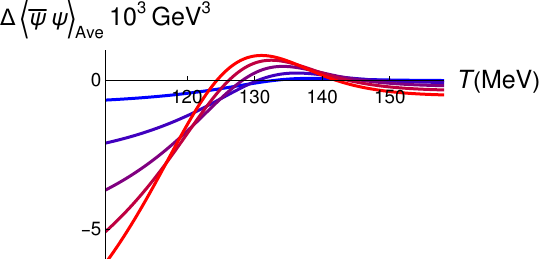}
  \caption{Condensate average (left panel) and condensate-average difference (right panel) as a function of temperature for different values of $eB$ with increasing strength from blue to red for $m_{\pi}=180$ MeV. The legends carry the same meaning as for the respective plots in Figs.~\ref{fig:cond_ave} and~\ref{fig:dcond_avedB}.}
  \label{fig:cond_ave_mpi180}
\end{figure}

Here, we display the results in a $2$-flavor nonlocal model for pion mass value, $180$ MeV (Fig.~\ref{fig:cond_ave_mpi180}). The IMC effect is seen for this pion mass. In the left panel of the figure, we have the condensate average capturing the IMC effect. The right panel, by showing the condensate-average difference, demonstrates clearly. The gray dashed line is the zero magnetic field crossover temperature at $m_\pi=180$ MeV. This is drawn to illustrate the temperature range where the IMC effect appears.  

\section{IMC effect for the individual light quark}
\label{sec:app_imc_indv}
\begin{figure}[!hbt]
\hspace{-0.8cm}
    \centering    
    \begin{subfigure}[b]{0.28\textwidth}
        \centering
        \includegraphics[width=\textwidth]{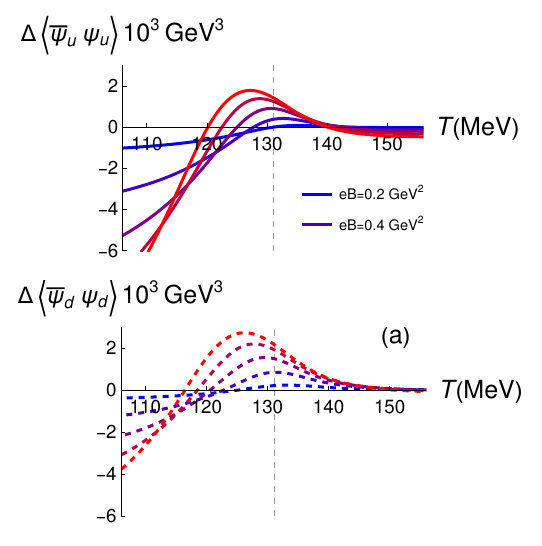}
        \caption*{}
        \label{fig:1}
    \end{subfigure}
%    \hfill
    \hspace{-0.8cm}
    \begin{subfigure}[b]{0.28\textwidth}
        \centering
        \includegraphics[width=\textwidth]{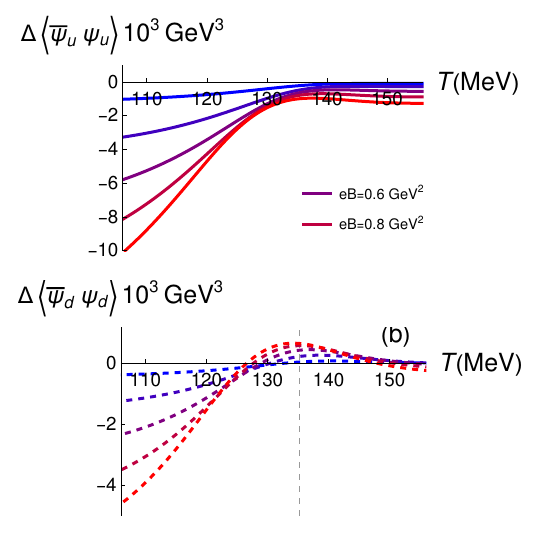}
        \caption*{}
        \label{fig:2}
    \end{subfigure}
%    \hfill
     \hspace{-0.8cm}
    \begin{subfigure}[b]{0.28\textwidth}
        \centering
        \includegraphics[width=\textwidth]{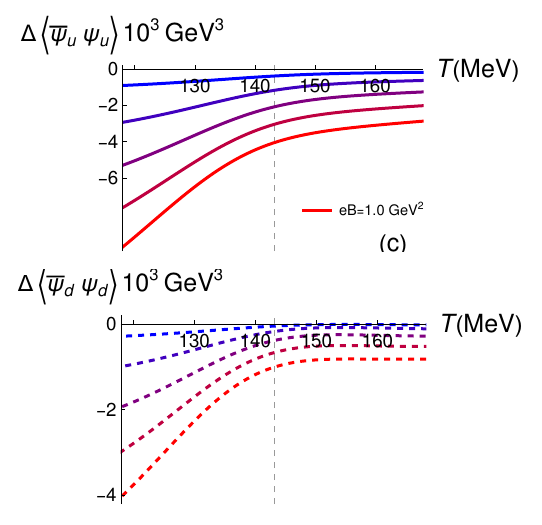}
        \caption*{}
        \label{fig:3}
    \end{subfigure}
%    \hfill
    \hspace{-0.8cm}
    \begin{subfigure}[b]{0.28\textwidth}
        \centering
        \includegraphics[width=\textwidth]{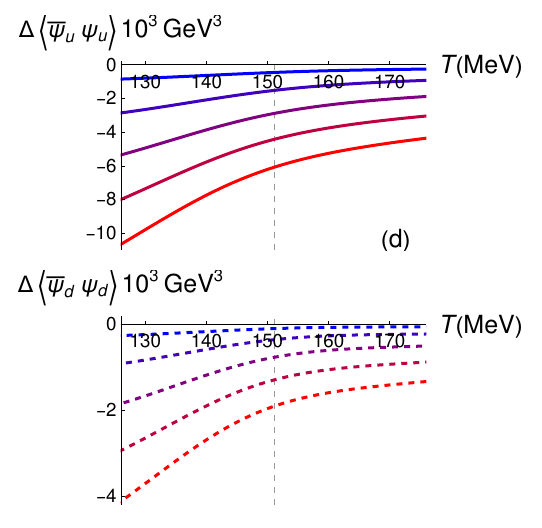}
        \caption*{}
        \label{fig:4}
    \end{subfigure}
    \caption{(Nonlocal): The condensate-difference for $u$ (solid) and $d$ (dashed) quarks are plotted as a function of temperature for different values of $eB$ with increasing strength from blue to red as denoted by the legends. The notations and legends are the same as in Fig.~\ref{fig:dcond_avedB}. The gray dashed lines represent the $T_{\rm CO}$ in the respective model at $eB=0$.}
    \label{fig:dcond_indv_nonlocal}
\end{figure}
In Fig.~\ref{fig:dcond_indv_nonlocal}, we have plotted the condensate difference for $u$ (top row of panels) and $d$ quarks (bottom row of panels) in a nonlocal NJL model. This is done to investigate the IMC effect for the individual light quarks. The definition for the individual quark flavors is similar to the definition for the condensate-average difference given in Eq.~\eqref{eq:cond_ave_diff}. For each flavor we define,
\begin{align}
\Delta\langle\bar\psi_f\psi_f\rangle(eB,T,m)=\langle\bar\psi_f\psi_f\rangle(eB,T,m)-\langle\bar\psi_f\psi_f\rangle(0,T,m).
\label{eq:cond_diff}
\end{align}

For a fixed $eB$ value, we observe from the figure (Fig.~\ref{fig:dcond_indv_nonlocal}) that with increasing pion mass the IMC effect for $u$ quark disappears at lower $m_\pi$ as compared to the $d$ quark. This might appear counter-intuitive as one expects the $u$ quark to get a stronger response (whether it is IMC or MC effect) with an increasing magnetic field than the $d$ quark due to the former's larger electric charge. Such an expectation should hold for all values of the pion mass, as we take the current quark mass to be equal for both the light flavors. However, this intuition holds true only when there is no mixing or coupling between different flavors~\cite{Ali:2020jsy}, or $G_{2}=0$ in Eq.~\eqref{eq:lag_each_term}, when the gap equations for the $u$ and $d$ flavors decouple. For $G_1=G_2$ (the usual form of the NJL model, the case we consider here), the gap equations for the $u$ and $d$ quarks are coupled [see Eq.~\eqref{eq:eff_mass_B}] since $\bar{\sigma}$ depends on both the flavors [Eq.~\eqref{eq:sigma}]. The second point to note is that mixing only affects the first term on the right-hand side of Eq.~\eqref{eq:cond_B_reg} --- the ``free'', ``free,reg'' parts of the condensate [Eq.~\eqref{eq:cond_B_free_reg}] are independent of $\sigma$. Since the mixing reduces the difference between the $u$ and $d$ condensates, it tends to reduce the IMC effect for $u$ but increases that for $d$. A larger contribution from the ``free,reg'' part for the $u$ condensate reduces the IMC effect further around the crossover region. On the other hand, with a smaller contribution from the ``free,reg'' term, the IMC persists for $d$ condensate up to a higher value of $m_{\pi}$ than that for the $u$ quark. However, for both flavors, the IMC effect disappears within the range of the studied pion mass. 

For fixed light $m_\pi$ values, note that the maximum value of $\Delta\langle\bar{\psi}_f\psi_f\rangle$ for each flavor increases monotonically with $eB$ signifying a stronger IMC effect near the crossover. For high $m_\pi$, $\Delta\langle\bar{\psi}_f\psi_f\rangle$ becomes more negative monotonically with increasing $eB$. Intuitively, one would expect such a regular pattern of the individual condensate in the presence of $eB$. 

\begin{figure}[!hbt]
\hspace{-1cm}
    \centering    
    \begin{subfigure}[b]{0.305\textwidth}
        \centering
        \includegraphics[width=\textwidth]{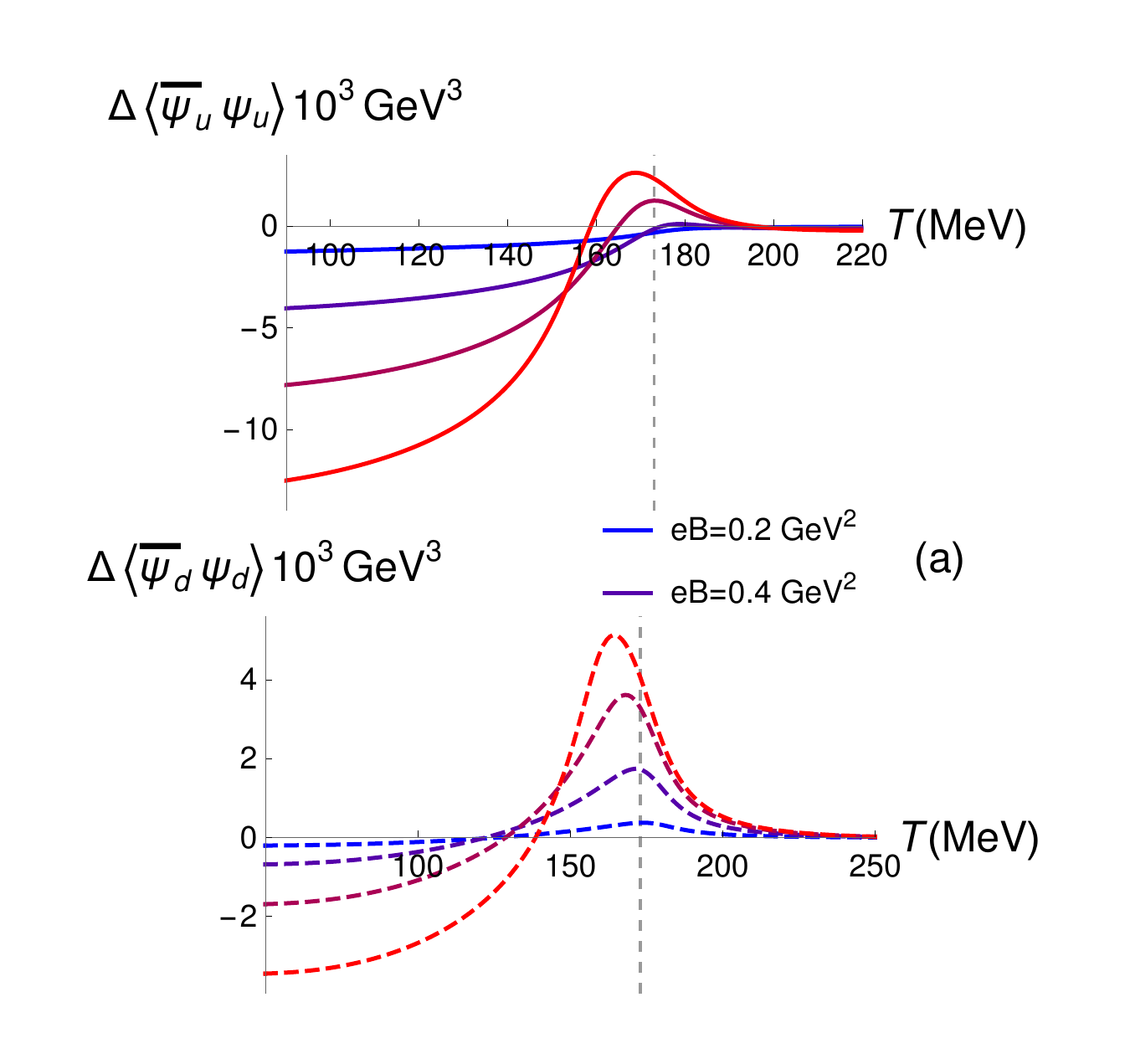}
        \caption*{}
        \label{fig:1}
    \end{subfigure}
%    \hfill
    \hspace{-1cm}
    \begin{subfigure}[b]{0.3\textwidth}
        \centering
        \includegraphics[width=\textwidth]{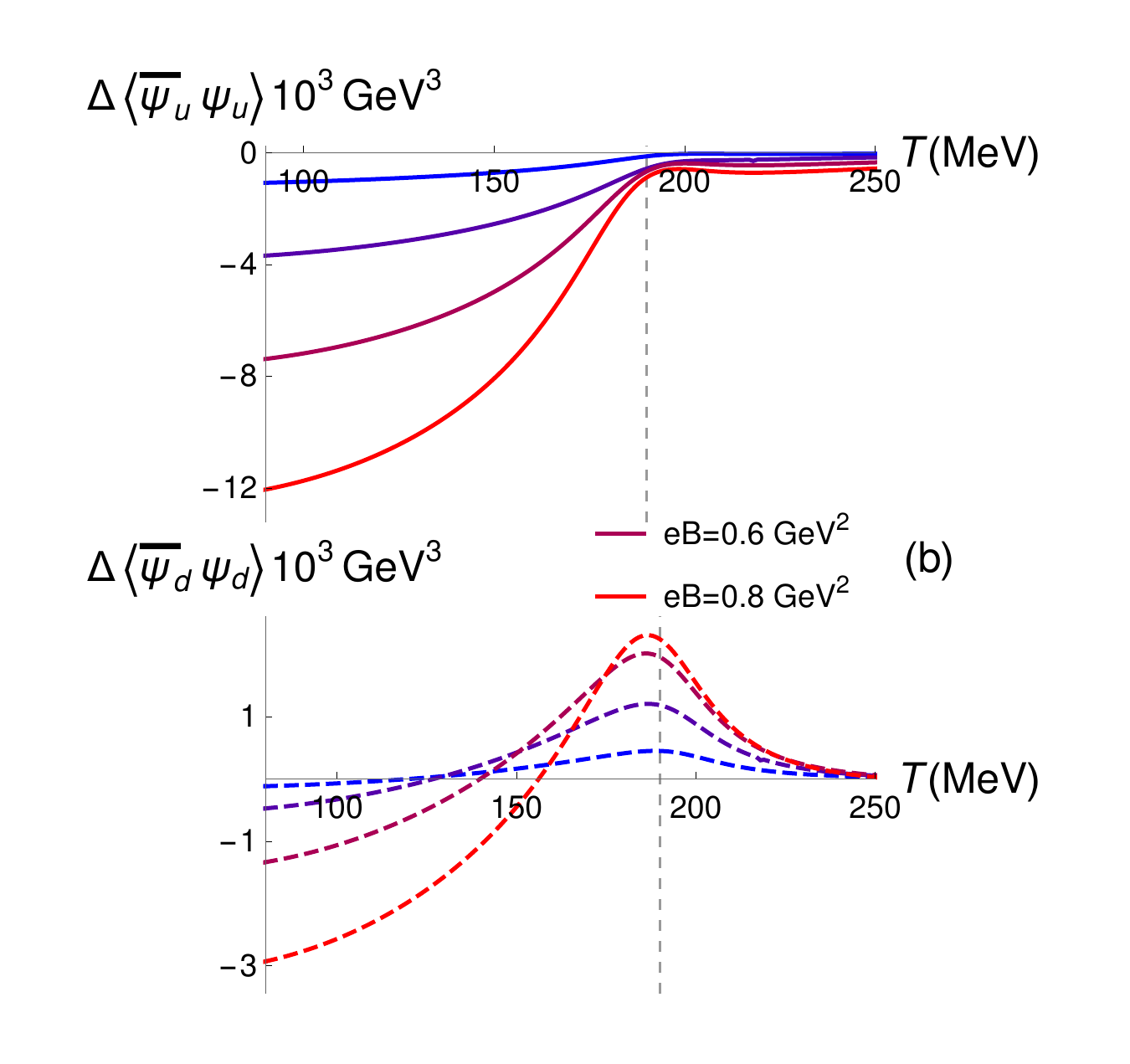}
        \caption*{}
        \label{fig:2}
    \end{subfigure}
%    \hfill
     \hspace{-0.8cm}
    \begin{subfigure}[b]{0.26\textwidth}
        \centering
        \includegraphics[width=\textwidth]{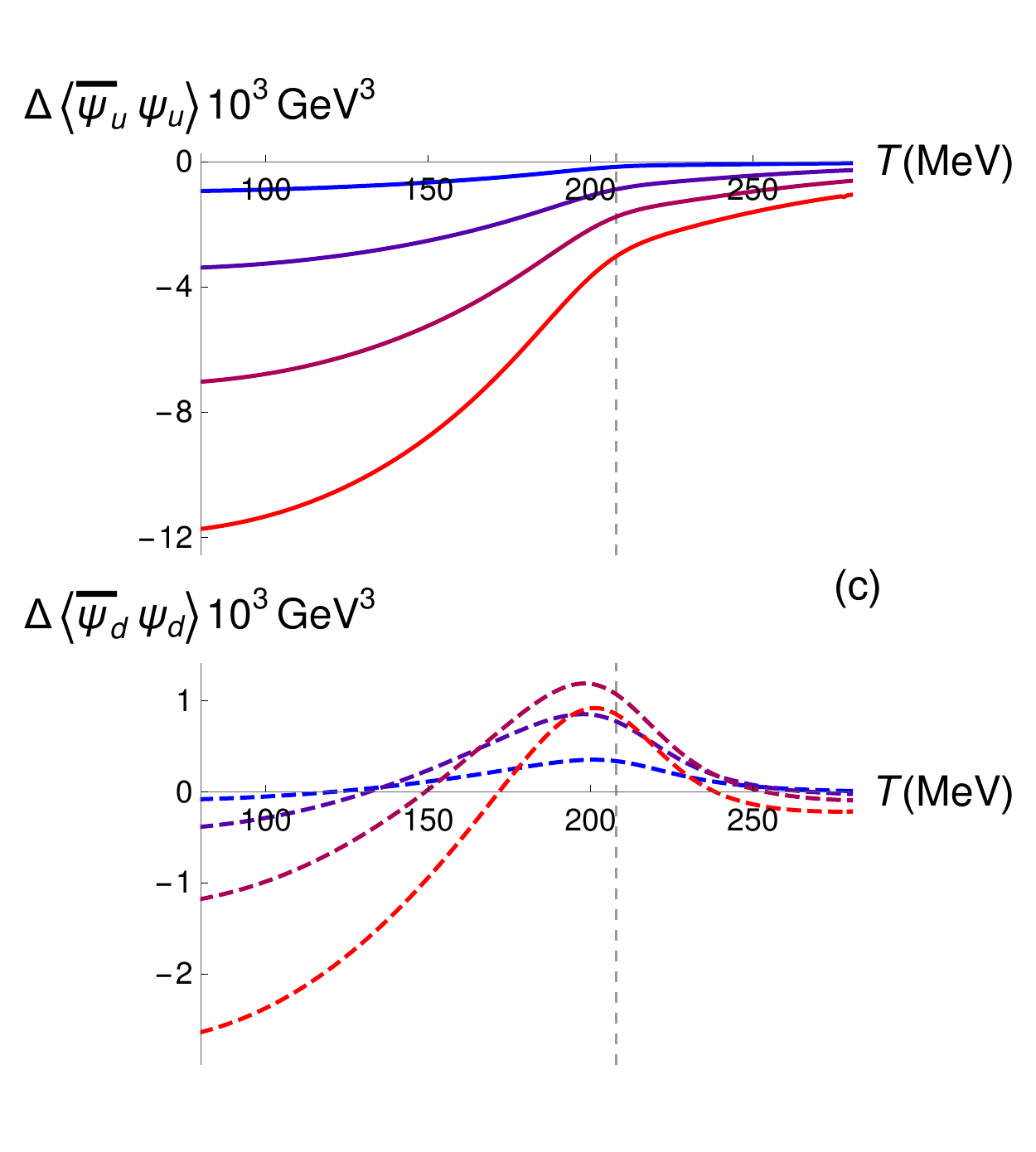}
        \caption*{}
        \label{fig:3}
    \end{subfigure}
%    \hfill
    \hspace{-0.8cm}
    \begin{subfigure}[b]{0.26\textwidth}
        \centering
        \includegraphics[width=\textwidth]{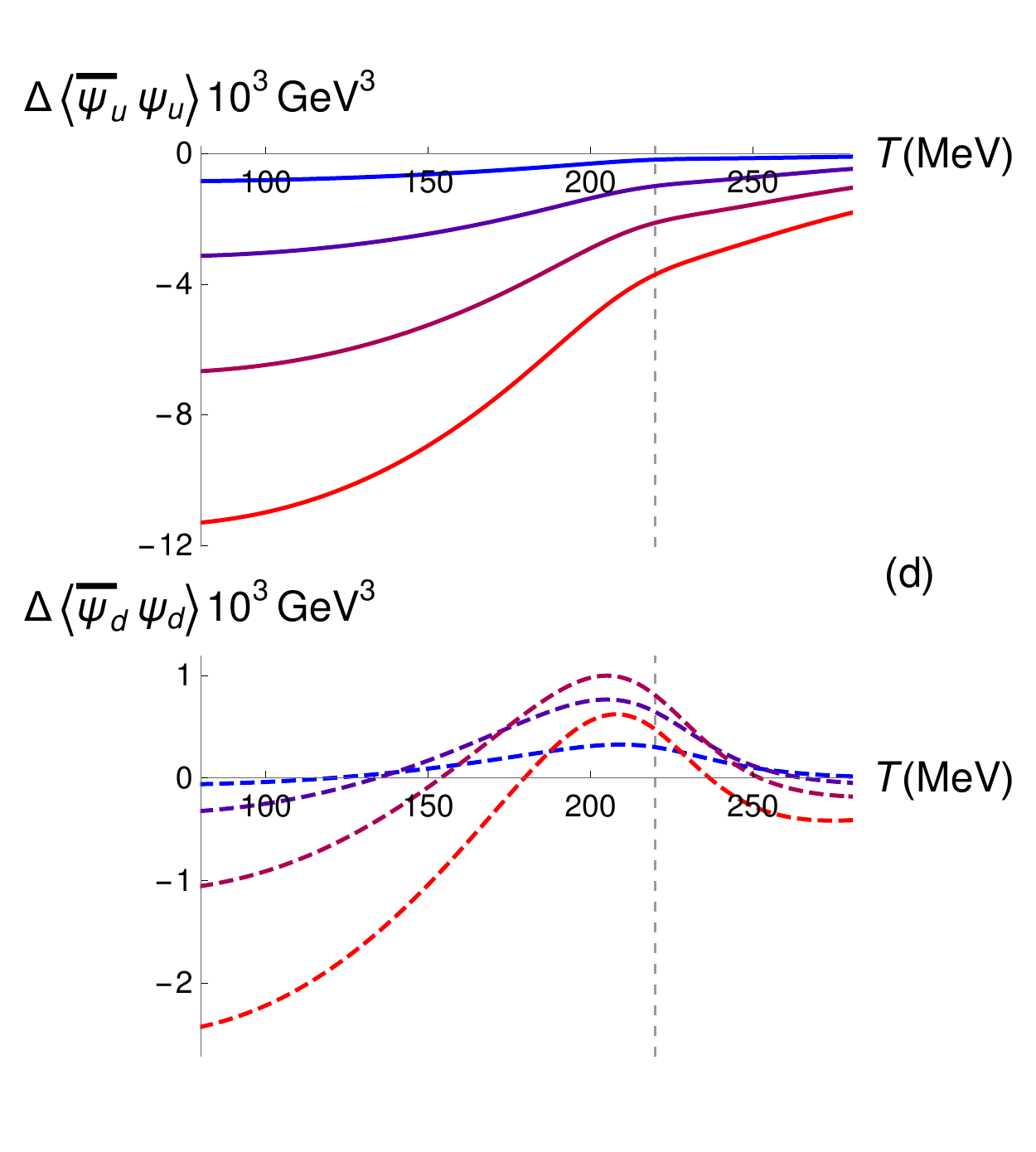}
        \caption*{}
        \label{fig:4}
    \end{subfigure}
    \caption{(Local $2+1$-flavor): The condensate-difference for $u$ (solid) and $d$ (dashed) quarks are plotted as a function of temperature for different values of $eB$ with increasing strength from blue to red as denoted by the legends. The notations and legends are the same as in Fig.~\ref{fig:dcond_avedB}. The gray dashed lines represent the $T_{\rm CO}$ in the respective model at $eB=0$.}
    \label{fig:dcond_indv_local2plus1f}
\end{figure}
The same plots in the $2+1$-flavor local model are shown in Fig.~\ref{fig:dcond_indv_local2plus1f}. In this case, with the increase of the pion mass, the IMC effect disappears for the $u$-quark, and the condensate behaves similar to the case in the nonlocal model with increasing $eB$-values. On the other hand, the IMC effect never disappears for the $d$ quark and survives for all the pion mass values that we tested. This is contrary to what we observed and explained in the nonlocal model. Also, the IMC effect is non-monotonic in $eB$ for the $d$ quark for fixed higher pion masses for different $eB$-values, as shown in panels (c) and (d). This irregular pattern might arise due to the flavor-independent running of the coupling as shown in the right panel of Fig.~\ref{fig:GvseB}. For the moment, no beyond-physical-point studies exist on individual light quark condensates or the condensate difference in LQCD calculations~\cite{DElia:2018xwo,Endrodi:2019zrl}. The models' predictions for individual quark condensates can be tested qualitatively once such data become available in the future.

\end{appendices}

\bibliography{ref_common}

\end{document}